\documentclass{aastex}	
\usepackage{emulateapj5}

\newcommand {\UIT}   	{{\em UIT\/}}
\newcommand {\Msun}	{\mbox{${\cal M}_\odot$}}
\newcommand {\Mbol}    {\mbox{$M_{\rm bol}$}}
\newcommand {\xten}[1]	{\mbox{$\times 10^{#1}$}}
\newcommand {\Teff}     {\mbox{$T_{\rm eff}$}}
\newcommand {\specline}[3] {\ion{#1}{#2}~$\lambda$#3}

\journalinfo{The Astronomical Journal, in press (February 2001)}
\slugcomment{ }

\shorttitle{UV and Optical Observations in Southwest Regions of the LMC}
\shortauthors{Parker, et al.}

\begin{document}

\title{Ultraviolet and Optical Observations of OB Associations and Field
Stars in the Southwest Region of the Large Magellanic Cloud}

\author{Joel~Wm.~Parker}
\affil{Southwest Research Institute, Suite 426, 1050 Walnut Street, Boulder,
Colorado 80302, USA}
\email{joel@boulder.swri.edu}

\author{Dennis Zaritsky}
\affil{Steward Observatory, Univ. of Arizona, 933 N. Cherry Ave,
Tucson, AZ 85721, USA}
\email{dzaritsky@as.arizona.edu}

\author{Theodore~P.~Stecher}
\affil{Laboratory for Astronomy and Solar Physics, Code 681, GSFC/NASA,
Greenbelt, Maryland 20771, USA}
\email{stecher@uit.gsfc.nasa.gov}

\author{Jason Harris}
\affil{UCO/Lick Observatory and Board of Astronomy and Astrophysics, University
of California, Santa Cruz, California 95064, USA}
\email{jharris@ucolick.org}

\and

\author{Phil Massey}
\affil{Lowell Observatory, 1400 W. Mars Hill, Flagstaff, AZ, 86001, USA}
\email{massey@lowell.edu}

\begin{abstract}

   Using ultraviolet photometry from the {\em Ultraviolet Imaging Telescope\/}
   (\UIT) combined with photometry and spectroscopy from three ground-based
   optical datasets we have analyzed the stellar content of OB associations and
   field areas in and around the regions N~79, N~81, N~83, and N~94 in the
   Large Magellanic Cloud.  In particular, we compare data for the OB
   association Lucke-Hodge~2 (LH~2) to determine how strongly the initial mass
   function (IMF) may depend on different photometric reductions and
   calibrations.  Although the datasets exhibit median photometric differences
   of up to 30\%, the resulting {\em uncorrected\/} IMFs are reasonably
   similar, typically $\Gamma \sim -1.6$ in the 5--60~\Msun\ mass range.
   However, when we correct for the background contribution of field stars, the
   calculated IMF flattens to $\Gamma = -1.3 \pm 0.2$ (similar to the Salpeter
   IMF slope).  This change underlines the importance of correcting for field
   star contamination in determinations of the IMF of star formation regions.
   It is possible that even in the case of an universal IMF, the variability of
   the density of background stars could be the dominant factor creating the
   differences between calculated IMFs for OB associations.

   We have also combined the \UIT\ data with the most extensive of these
   ground-based optical datasets --- the Magellanic Cloud Photometric Survey
   --- to study the distribution of the candidate O-type stars in the field.
   We find a significant fraction, roughly half, of the candidate O-type stars
   are found in field regions, far from any obvious OB associations [in accord
   with the suggestions of \citet{GCC82} for O-type stars in the solar
   neighborhood].  These stars are greater than 2~arcmin (30~pc) from the
   boundaries of existing OB associations in the region, which is a distance
   greater than most O-type stars with typical dispersion velocities will
   travel in their lifetimes.  The origin of these massive field stars (either
   as runaways, members of low-density star-forming regions, or examples of
   isolated massive star formation) will have to be determined by further
   observations and analysis.

\end{abstract}




\section{INTRODUCTION}

The most fundamental characterization of star formation is the slope of the
initial mass function (IMF); it is a crucial parameter in our theoretical
understanding of astrophysical topics from star formation \citep{R94, E97} to
galaxy evolution \citep[e.g.,][]{Letal96}.  In a common notation, $\Gamma$ is
the index of the power law used to define the IMF (i.e., the slope in the
log-log plot of number versus mass distribution).  This parameter is usually
assumed to be universally near the \citet{S55} value ($\Gamma=-1.3$).

However, evidence now suggests that the IMF depends strongly on some properties
of the star formation environment.  \citet{MLDG95} and \citet{Petal96, Petal98}
identify a significant {\em field\/} component of massive stars in the Clouds.
Massey~et~al. found an extremely steep field IMF ($\Gamma \sim -4$), but this
result depended upon a large incompleteness correction determined from two
small areas (the incompleteness correction was significant for the fainter,
typically lower mass stars, with likely minimal effect for the high mass stars;
for this reason, their IMF results heavily depend on the highest mass bins).
Parker~et~al. found a flatter IMF, but relied purely upon ultraviolet (UV)
photometry rather than optical spectra to determine temperatures, bolometric
corrections, and hence masses.  A spectroscopic study of the Magellanic Cloud
field stars is now underway to address the discrepant results for the IMF.

A physical framework that may explain many aspects of the observed IMF is
provided by random sampling of fractal clouds, which naturally produces a
Salpeter IMF for all sizes of clouds \citep{E97, E99a, E00, MS00}.  According
to that model, in well-sampled regions formed from large clouds (such as OB
associations), one would measure a Salpeter IMF.  However, field regions could
display a steeper observed IMF due to: a superposition of undersampled Salpeter
IMFs from many star-forming clouds with a large range of sizes, differential
drift as a function of mass, or lower local ISM and cloud pressure allowing
more efficient cloud disruption.  In particular, as described by \citet{E99b},
if the higher mass stars can sometimes destroy clouds and halt further local
star formation, then even if all clouds sample the Salpeter IMF, low mass
clouds will make primarily low mass stars, and high mass clouds will make low
and high mass stars.  If there are many more low mass clouds than high mass
clouds (such as may be the case in the distant field regions), then the
composite IMF from all these clouds will be measured to have a slope steeper
than the Salpeter slope, i.e., have proportionately more lower mass stars
because of the preponderance of low mass clouds.  \citet{EH00} find that the
general field pressure in dwarf irregular galaxies is roughly an order of
magnitude lower than the pressure in the H\,{\sc ii} regions, unlike the case
in spiral galaxies where the two pressures are roughly equal.  This difference
could explain why the IMF in the LMC field could be relatively steep, whereas
the IMF in the Galactic field and in Galactic and LMC OB associations could be
similar to the Salpeter IMF.  \citet{MLDG95} note that their analysis of the
\citet{GCC82} data implies a steep IMF slope for Galactic field stars, which
would contradict the theoretical model, but they also point out the data may be
strongly biased by selection effects.

Much of the uncertainty in the field star IMF may be due to our lack of
knowledge of the population of massive OB-type stars in the field.  In an
analysis of the N~11 region in the northwest LMC, \citet{Petal96} discuss the
O-type star content in the OB associations and in the nearby field regions.
They used UV photometry from the {\em Ultraviolet Imaging Telescope\/} (\UIT)
for the full region and environs; $UBV$ photometry for the OB associations
LH~9, 10, and 13 (but not for the field stars) was available in the literature
for multi-wavelength analysis.  From the combination of UV and optical
ground-based photometry they identified 88 candidate O-type stars (COTS) in the
LH~9, 10, and 13 fields (in a total area of $\sim 41$~arcmin$^{2}$) and as many
as 170 to 240 such stars in the entire 37~arcmin-diameter field-of-view.
However, this estimate for the population of COTS depended solely on UV data in
a single filter, requiring various assumptions for the O and B star fraction in
the field and the range of reddening.

In this paper, we take the next step in determining the population of massive
OB-type stars throughout the Clouds and understanding the IMF in the field and
associations.  We analyze the COTS population for another LMC region, but this
time using both UV and optical data not only for stars in OB associations, but
for field stars as well.  Although we cannot yet determine the IMF for the
field stars due to lack of spectroscopic classifications of the COTS (a project
that, as mentioned above, is currently underway), we do have spectra for the
blue stars in one of the OB associations in the region, and so we analyze the
IMF for that association.  In Section~\ref{sec:Data} we discuss the region
observed and the various datasets used in this study.  In
Section~\ref{sec:Comps}, we make cross-comparisons between three optical
datasets:  the Magellanic Cloud Photometric Survey \citep[MCPS;][]{Zetal97},
the catalog of \citet{O96}, and heretofore unpublished photometry.  We analyze
these datasets in a consistent manner and compare the resulting IMFs for the OB
association to better understand the potential differences that may arise in
IMF studies.  In Section~\ref{sec:UV} we combine the UV and optical datasets to
calculate temperatures of the stars and to analyze the population of hot stars
throughout the region, and better determine the distribution of COTS in the
field, far from currently known OB associations.


\section{DATASETS} \label{sec:Data}

\subsection{\UIT\ UV Photometry}

\begin{figure*}

\plotone{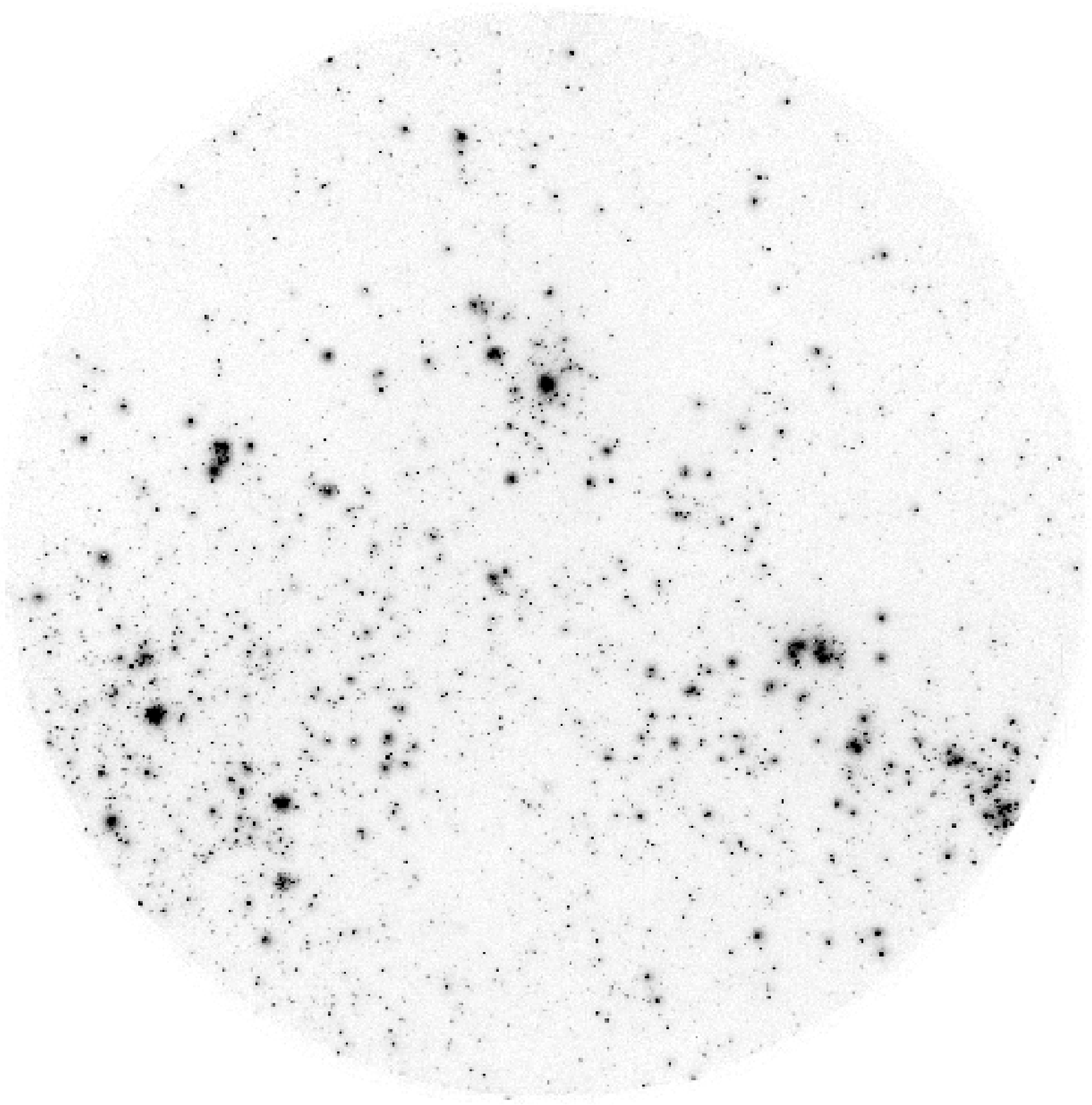}
\plotone{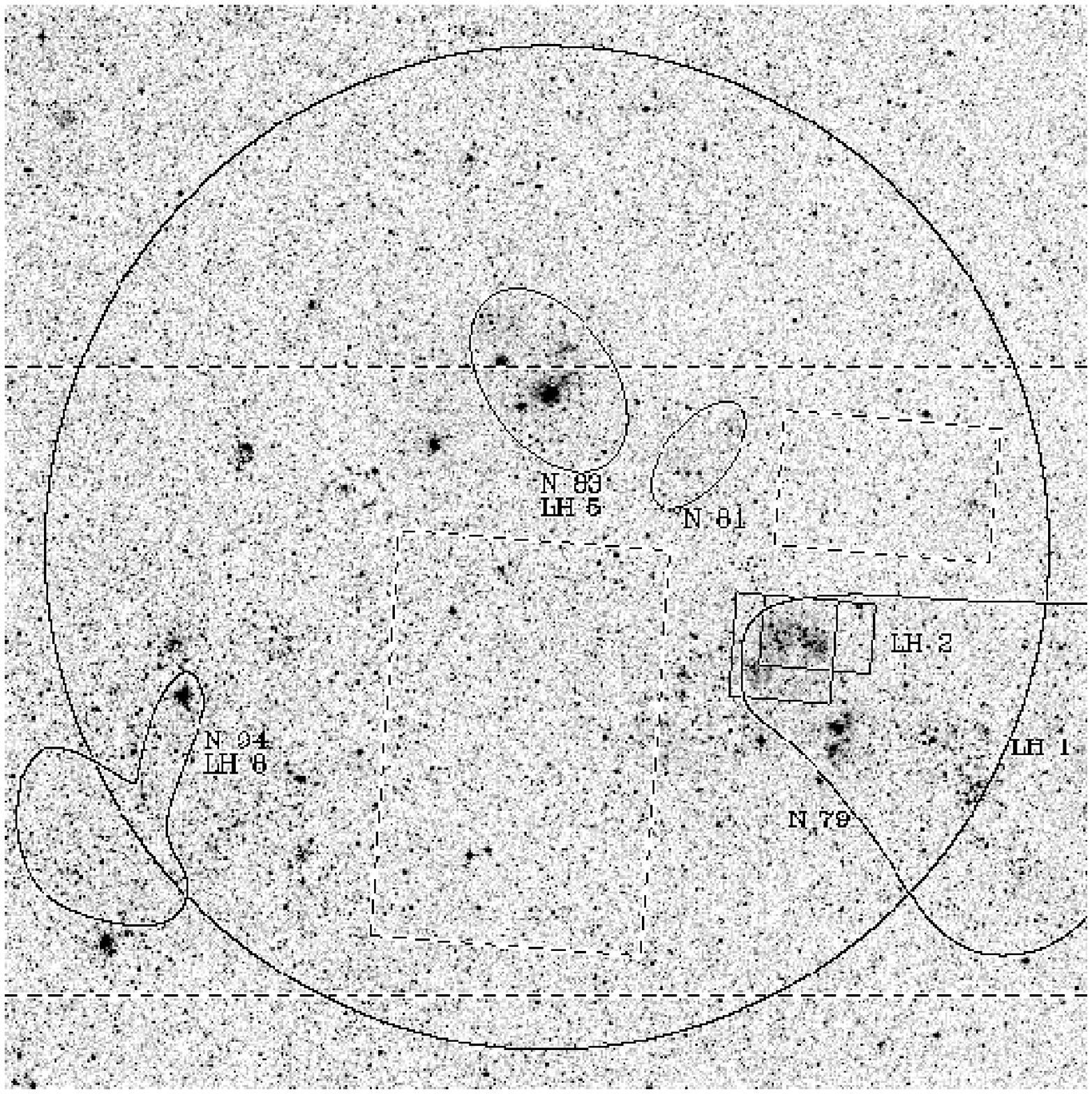}

\caption[]{a) The \UIT\ image used in this analysis.  b) The the corresponding
image from the Digitized Sky Survey.  Both images are at the same orientation
(north is about 4 degrees clockwise from up, east is to the left) and scale
(the \UIT\ image is roughly 37~arcmin in diameter).  Several regions are
outlined on the Digitized Sky Survey image: the large circle shows the position
of the \UIT\ image; the two long, horizontal dashed lines delineate the limits
of the field covered by the MCPS; the elliptical and irregular solid outlines
indicate Henize~(1956) regions; the solid-lined small boxes to the west show
the orientation of the CCD images from CTIO85 (rectangular box) and from
\citet{O96} (square box) that define the area used in the multi-dataset
analysis of the OB association LH~2; and the two larger dash-lined rectangles
in the west and the central-south show the ``field'' regions used to estimate
the background field star corrections to the IMF determined from the MCPS
data.}
\label{fig:uit_dss}

\end{figure*}

Figure~\ref{fig:uit_dss}a shows the \UIT\ image used in this analysis, and
Figure~\ref{fig:uit_dss}b shows the corresponding image from the Digitized Sky
Survey with various OB associations, clusters, and ground-based CCD fields
identified.  The \UIT\ field covers parts of N~79 and N~94, all of N~81, N~83,
and a number of smaller regions.  The identification of these regions
(LHa~120-N~79, 81, 83, 94) was originally defined by \citet{H56} in his catalog
of H$\alpha$-emission stars and nebulae.  The extent of those regions are also
shown in the LMC atlas by \citet{HW77}, and are outlined in
Figure~\ref{fig:uit_dss}.  Some Lucke-Hodge \citep[LH;][]{LH70} regions are
also identified in Figure~\ref{fig:uit_dss}b.  N~79 is an irregular region in
the southwest part of the \UIT\ image.  It is roughly $17 \times 14$~arcmin in
size, and encloses the regions NGC~1712 (LH~1, which is just within the edge of
the \UIT\ image), 1722, and 1727 (LH~2).  N~83 near the center of the region is
about 5~arcmin in diameter, and contains the OB association LH~5, which is
comprised of NGC~1737, 1743, 1745, and 1748.  To the southeast, N~94 is
identified as LH~8, and encloses NGC~1767.


The details of the \UIT\ catalog (instrument, observations, and data reduction)
are fully discussed by \citet{Setal92, Setal97} and \citet{Petal98}.  During
the Spacelab Astro-2 mission which flew aboard the space shuttle Endeavour on
1995 March 2-18, \UIT\ obtained more than 700 ultraviolet images of nearly 200
celestial targets.  The images include 16 fields in the LMC and three fields in
the SMC \citep{Petal98}.  The \UIT\ observations of the field in
Figure~\ref{fig:uit_dss}, designated as ``N~79'' in the catalog of
\citet{Petal98}, were made on 1995 March 14, and consist of two exposures
(65~sec, 653~sec) in the B5 filter, which has a centroid wavelength of $\lambda
= 1615$~\AA, and a bandwidth of $\Delta \lambda = 225$~\AA\ [see
\citet{Setal92} for the filter response curve].  The $\sim$37~arcmin diameter
photographic images were scanned and digitized with a PDS 1010m
microdensitometer, resulting in images with 1.13~arcsec pixels and point-source
profiles with FWHM$= 3.36 \pm 0.29$ arcsec.  Calibrations were made in the same
manner as for Astro-1 \citep{Setal92}, based on laboratory measurements and
data obtained during the missions.  Flux value zeropoints were derived
primarily with comparisons to {\em IUE\/} stars, but also with comparisons to
stars observed by {\em OAO-2\/}, {\em HUT\/}, {\em ANS\/}, {\em GHRS\/}, and
other UV-capable instruments.  UV magnitudes are defined from these fluxes as:
$m_{\rm UV} = -2.5 \log(F_{\lambda}) - 21.1.$ Astrometry was performed with
reference to {\em HST\/} guide stars \citep{Letal90}.

Stellar photometry on the \UIT\ images was performed with {\sc idl} procedures
based on the {\sc daophot} algorithms \citep{S87}.  Aperture corrections were
calculated for each image, and small corrections were made in the zeropoint
offsets so that the median difference of the flux for all stars was zero
between the two images (putting both images on the same zeropoint).  The final
magnitude for each star is the average of its measurements on the two images
weighted by the inverse square of its calculated photometric errors.  A
comparison with {\em IUE\/} observations of three relatively uncrowded stars in
the field show that the \UIT\ and {\em IUE\/} fluxes agree to better than 5\%.
The completeness limits of the ground-based data used in this paper go to
slightly later spectral types, so the combined \UIT\ and ground-based dataset
is limited by the UV data.

\begin{figure*}

\epsscale{1.5}
\plotone{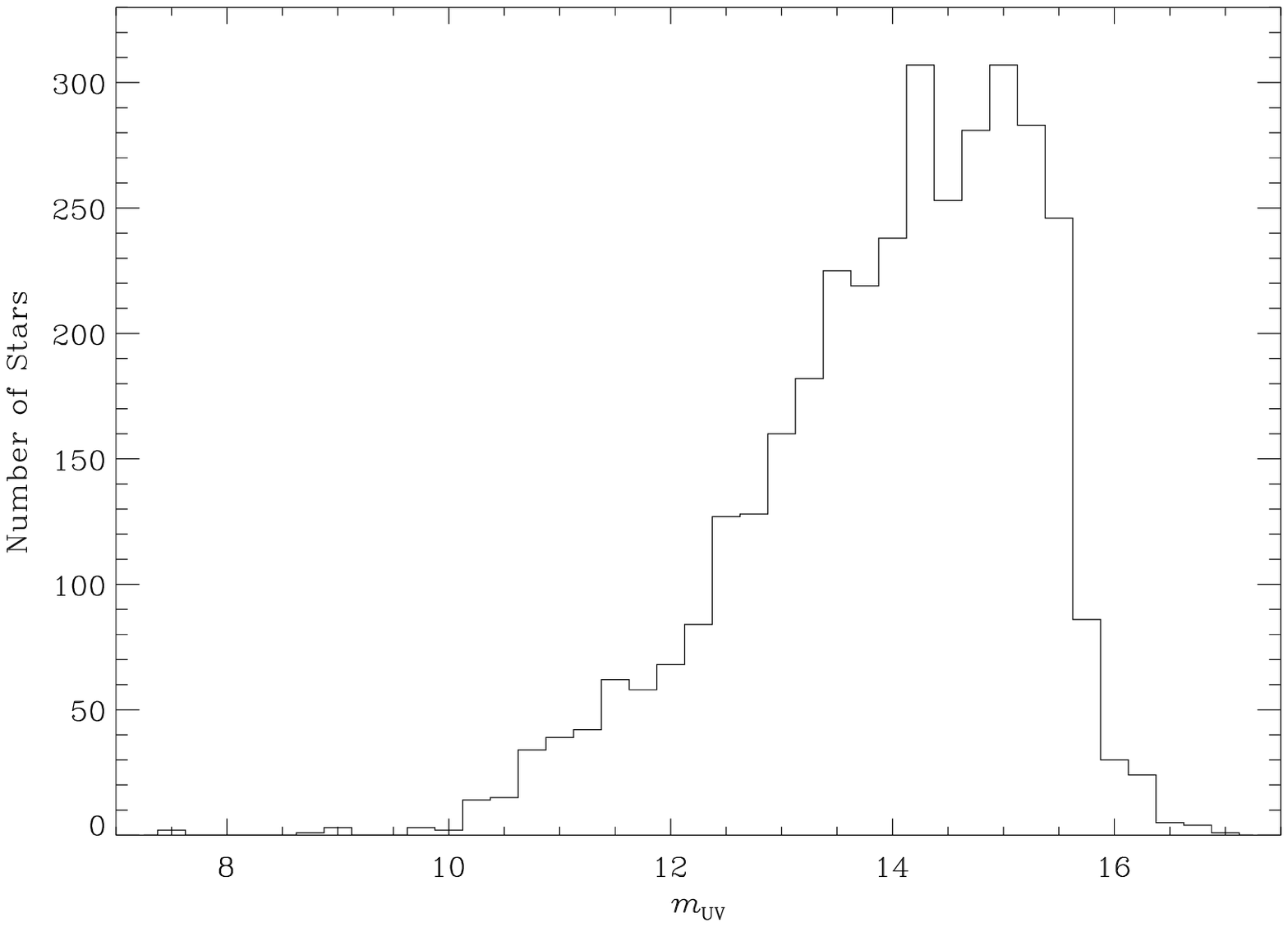}

\caption[]{The UV ($\lambda = 1615$~\AA, $\Delta \lambda = 225$~\AA) luminosity
function for all the stars detected in the \UIT\ field
(Figure~\protect\ref{fig:uit_dss}a).}
\label{fig:lf_uit}

\end{figure*}

Figure~\ref{fig:lf_uit} shows the histogram of observed UV magnitudes for the
3533 stars detected in the \UIT\ images of this field.  The limiting magnitude
is $m_{\rm UV} \approx 17$~mag, the magnitude of an unreddened late B-type
($\sim$~B9) star.  The completeness limit, the magnitude to which we should
have detected all stars, is $m_{\rm UV} \approx 15$~mag, the magnitude of an
unreddened early B-type ($\sim$~B3) star.


\begin{figure*}

\vspace*{4ex}
\epsscale{1.7}
\plotone{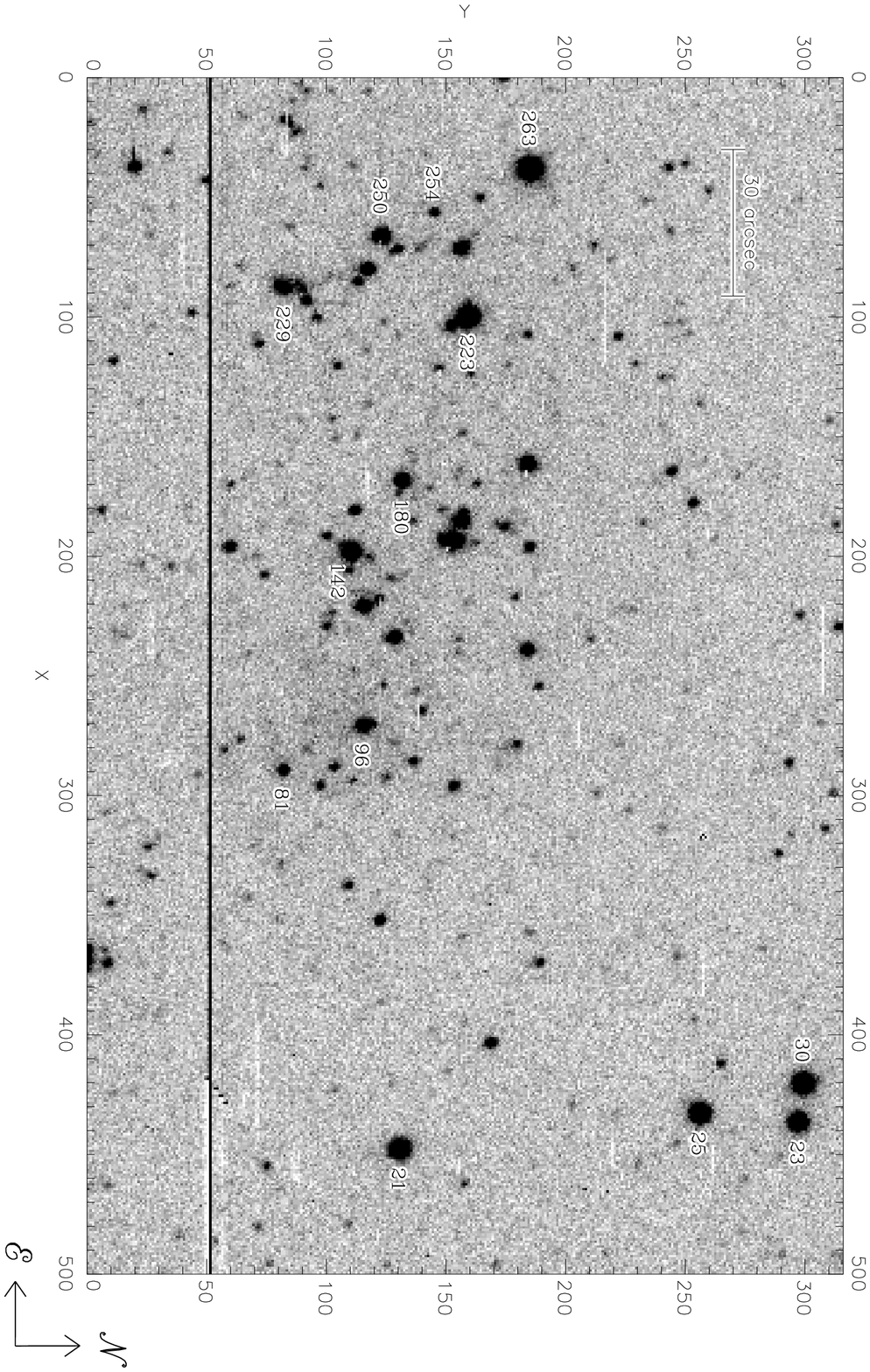}

\vspace*{8ex}
\caption[]{The CCD image from the CTIO85 data (c.f.,
Figure~\protect\ref{fig:uit_dss}b).  North and east are indicated by the
arrows.  Numbers identify those stars that were observed spectroscopically.}
\label{fig:ccd}

\end{figure*}

\subsection{Ground-based Optical Photometry}

The MCPS catalog is discussed by \citet{Zetal97}. The data originate from
$UBVI$ drift-scan imaging on the Las Campanas Swope (1~m) telescope with the
Great Circle Camera \citep{Zetal96} and a 2048$\times$2048 pixel CCD with pixel
scale of 0.7~arcsec~pixel$^{-1}$. Typical seeing is $\sim$1.5~arcsec.  The data
reduction and uncertainties are described in the original paper.  The survey is
still progressing, so the data presented here are drawn from a preliminary
catalog of a small region of the LMC, and for that reason, we do not publish
the MCPS data here.  (The \UIT\ catalog is already publically available.)  The
MCPS data for this and other regions will be made public when reductions of
nearby regions are complete, and may be superior only in that the overlapping
areas between all scans will enable checks and possible corrections to the
photometry.  As seen in Figure~\ref{fig:uit_dss}b, the currently available scan
region does not cover the entire \UIT\ field, so for our UV+optical analysis we
can use only this region of overlap.  The full scan of this region, obtained in
1997 December, is 124$\times$25 arcmin$^2$, resulting in a catalog of 265,794
stars; 80,733 stars are in the region that overlaps with the \UIT\ image.  The
area of this overlapping region is $\sim$675~arcmin$^2$, or
1.4\xten{5}~pc$^2$.

\begin{table*}
\caption{The CTIO85 Catalog\tablenotemark{a} : Photometry and Astrometry of
Stars in LH 2 \label{tab:phot}}
{\footnotesize
\begin{center}
\begin{tabular}{r rr rr rrrrrr ll}
\hline \hline
{Star} &
\multicolumn{2}{c}{$\alpha$ \ \ \ (2000.0) \ \ \ $\delta$} &
{$X$ \tablenotemark{b}} &
{$Y$ \tablenotemark{b}} &
{$V$} &
{$\sigma_V$} &
{$B-V$} &
{$\sigma_{B-V}$} &
{$U-B$} &
{$\sigma_{U-B}$} &
{O96 ID \tablenotemark{c}} &
{Spectral Type} \\
\hline
 & & & & & & $\vdots$ \\
81 & 4:52:00.66 & --69:20:49.73 & 288.80 &  81.71 & 15.704 &  0.014 &  0.029 &  0.024 & --0.745 &  0.031 & D10b-23 & O6 V \\
82 & 4:52:00.79 & --69:20:39.39 & 287.52 & 102.83 & 16.678 &  0.030 &  0.914 &  0.085 & --0.243 &  0.178 & D10b-44 \\
83 & 4:52:00.96 & --69:19:06.23 & 285.84 & 293.19 & 16.844 &  0.047 &  1.294 &  0.109 & 99.000 &  9.000 & D10b-55 \\
84 & 4:52:00.98 & --69:20:05.92 & 285.49 & 171.23 & 18.717 &  0.136 & --0.179 &  0.244 & 99.000 &  9.000 & D10b-308 \\
85 & 4:52:01.02 & --69:20:23.15 & 284.98 & 136.02 & 16.466 &  0.023 & --0.163 &  0.036 & --0.674 &  0.053 & D10b-39 \\
86 & 4:52:01.06 & --69:20:50.53 & 284.46 &  80.08 & 18.927 &  0.320 & --0.665 &  0.383 & 99.000 &  9.000 & D10b-504 \\
87 & 4:52:01.32 & --69:18:55.21 & 282.04 & 315.71 & 18.285 &  1.482 &  1.363 &  1.509 & 99.000 &  9.000 & D10b-500 \\
88 & 4:52:01.48 & --69:21:01.59 & 279.93 &  57.46 & 17.663 &  0.065 & --0.147 &  0.102 & --0.196 &  0.159 & D10b-118 \\
89 & 4:52:01.50 & --69:20:44.84 & 279.80 &  91.69 & 17.864 &  0.064 & --0.077 &  0.123 & --0.751 &  0.186 & D10b-187 \\
90 & 4:52:01.51 & --69:20:04.57 & 279.70 & 173.99 & 18.366 &  0.107 &  0.278 &  0.251 & --1.558 &  0.297 & D10b-203 \\
91 & 4:52:01.69 & --69:20:02.02 & 277.86 & 179.19 & 17.038 &  0.035 & --0.166 &  0.065 & --0.596 &  0.087 & D10b-68 \\
92 & 4:52:01.69 & --69:19:40.68 & 277.85 & 222.79 & 18.134 &  0.081 &  0.571 &  0.312 & 99.000 &  9.000 & D10b-297 \\
93 & 4:52:01.71 & --69:20:26.92 & 277.61 & 128.31 & 18.013 &  0.074 & --0.065 &  0.138 & --0.334 &  0.244 & D10b-181 \\
94 & 4:52:01.86 & --69:20:58.37 & 275.84 &  64.04 & 17.404 &  0.054 &  1.222 &  0.177 & 99.000 &  9.000 & D10b-105 \\
95 & 4:52:02.27 & --69:20:16.21 & 271.56 & 150.19 & 18.625 &  0.142 &  0.353 &  0.301 & --1.225 &  0.460 & D10b-435 \\
96 & 4:52:02.40 & --69:20:33.32 & 270.12 & 115.21 & 14.947 &  0.019 & --0.170 &  0.024 & --0.870 &  0.023 & D10b-12 & O7.5 Vz \\
 & & & & & & $\vdots$ \\
\hline \hline
\end{tabular}
\end{center}

\vspace*{-8ex}

\tablenotetext{a}{The complete version of this table is in the electronic
edition of the Journal.  The printed edition contains only a sample.}

\tablenotetext{b}{The $X$ and $Y$ coordinates correspond to the coordinate
system of the CCD image in Figure~\protect\ref{fig:ccd}}

\tablenotetext{c}{The ``O96 ID'' is the identification of the star in the
catalog of \citet{O96}.}

}

\end{table*}

In addition to these two primary datasets, we also use $UBV$ observations from
two other sources.  One source of $UBV$ data is from observations made by
P.~Massey and K.~DeGioia-Eastwood on 1985 February 12 with the 0.9~m telescope
at CTIO.  The observed LH~2 field covers an area of 4.1$\times$2.6~arcmin$^2$;
the $V$-band CCD image is shown in Figure~\ref{fig:ccd}.  Details of those
observations are given by \citet{MSGD89}. The data were reduced with {\sc
daophot} by J.Wm.P. in 1990, and used the transformations to the standard
system defined by \citet{MSGD89}.  The reduction of these LH~2 data was similar
to the reductions used in the analyses of other OB associations observed on the
same run \citep[e.g.,][]{MSGD89, MPG89, PGMW92, P93, GMP94, OM95}.  Astrometric
positions were calculated from the CCD X,Y coordinates fitted by {\sc daophot}
and using a plate solution derived with the Grant machine at the offices of the
National Optical Astronomical Observatories.  A total of 280 stars are in this
catalog for LH~2; these data (hereafter:  CTIO85) were discussed by
\citet{PGM90}, but had not been published, so for completeness we present them
here in Table~\ref{tab:phot}.

Another source of $UBV$ data is from the catalog of \citet{O96}; details of the
observations (made in 1992 November--December) and data reductions are
described therein. A total of 674 stars are in that catalog for LH~2, which
covers an area of 3.8$\times$3.8~arcmin$^2$.  The areas covered by the
\citet{O96} and CTIO85 datasets are shown by the rectangular solid outlines in
Figure~\ref{fig:uit_dss}b.

\subsection{Spectroscopy}

UBV photometry alone cannot adequately distinguish O-type from B-type stars
\citep{M85}; \citet{C99} provides an analysis of the reliability and
limitations of $UBV$ photometry to derive stellar effective temperatures
(\Teff).  Though UV photometry can help reduce this degeneracy, it still is not
as accurate as using spectral types to determine \Teff\ for hot stars.
\citet{M98} discusses the utility for UV photometry based on the
\citet{Hetal97} {\em HST\/} UV and optical observations of 30~Doradus.  He
shows that the inclusion of UV photometry can help better determine \Teff\ for
the later-type O stars, but for the hottest, early-type O stars a photometric
accuracy of 5\% translates to about 1~mag uncertainty in the bolometric
correction.

\begin{figure*}

\hspace*{-7.5in}
\plotfiddle{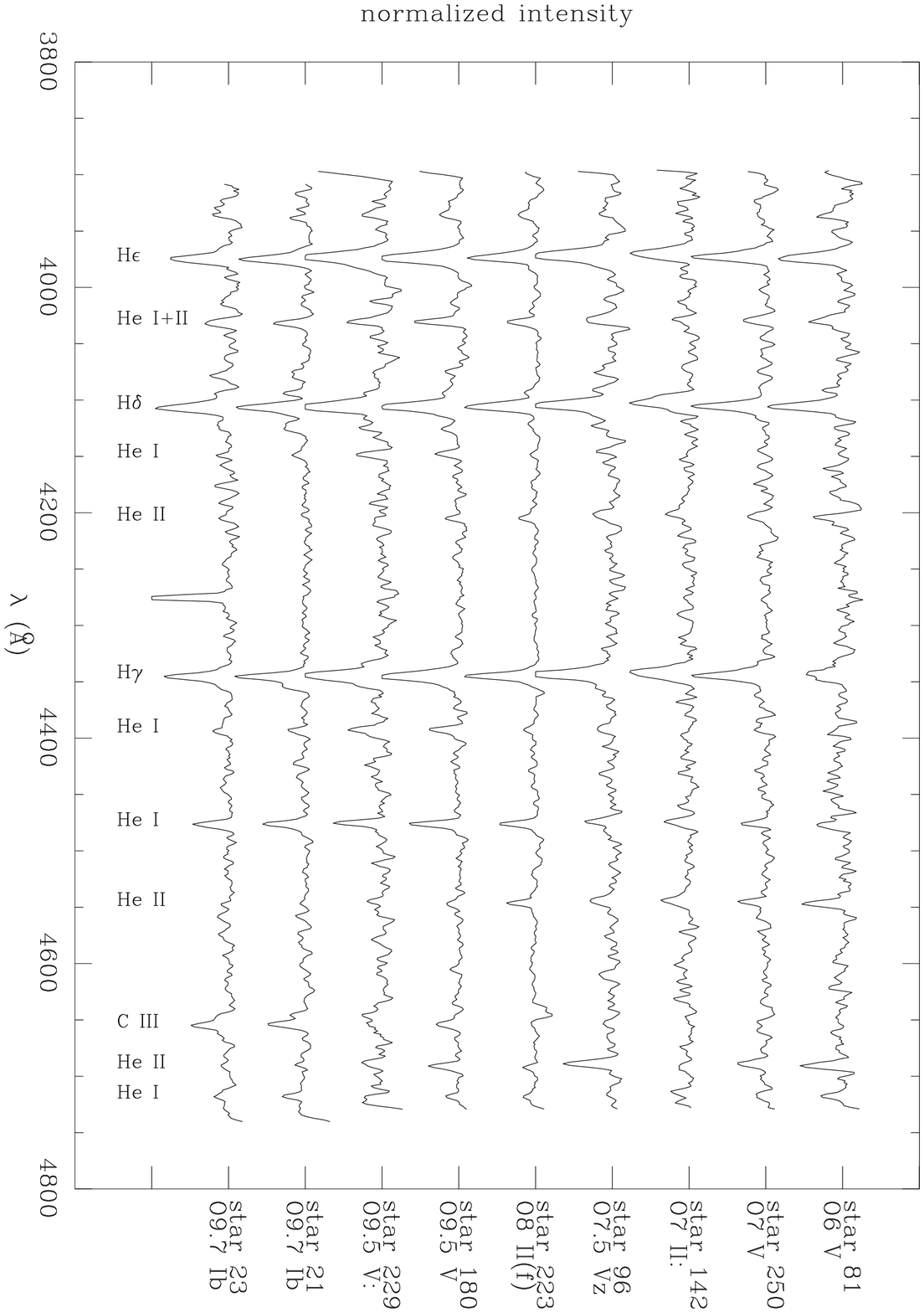}{3in}{180}{0.9}{0.9}{0}{0}

\caption[]{Normalized spectra of the early-type stars in LH~2.  Each tick mark
on the Y-axis represents 0.25 continuum units.}
\label{fig:spectra}

\end{figure*}

We therefore also have obtained spectroscopic classifications of stars in the
LH~2 OB association.  We observed the bluest [$Q = (U-B) - 0.72 \times (B-V) <
-0.7$] and brightest ($V < 16$~mag) stars in LH~2 based on the CTIO85
photometry, and also observed selected red stars to determine membership and
identify foreground stars.  Spectra for these stars were obtained on
1991~Jan~29 using the CTIO 4~m telescope and R/C spectrograph with a GEC CCD
and the Air Schmidt camera.  The KPGL1 grating (632 lines/mm blazed at
4200~\AA) was used in first order to obtain a wavelength coverage of
3910--4740~\AA.  The resolution was $\sim 2.7$~\AA\ (2.0 pixels at
1.3~\AA/pixel, which for 22~$\mu$m pixels gives 59~\AA/mm).  Exposure times
longer than about 10 minutes were divided into three separate exposures to
allow identification and removal of cosmic rays.  These observations were made
in conjunction with another observing program \citep{PGMW92, P93}, and further
details of the observations and reductions can be found therein.  The spectra
and classifications of the observed early-type stars are shown in
Figure~\ref{fig:spectra}.


Classifications were made by comparing observed spectra with: digital spectral
standards \citep{WF90}, standards taken during the run, and all stars observed
during the same run (including spectra of other LMC stars for another project)
to insure consistency.  Walborn (private communication) kindly provided an
independent check of the classifications.  The earliest type star in the LH~2
OB association is star \#81, which is classified as a main-sequence O6 star,
implying (along with isochrones fitted to the H-R diagrams discussed in the
next section) an approximate cluster age of $\lesssim 3$~Myr, assuming
coevality.  The nearby star \#96 is classified as a O7.5~Vz type; the ZAMS (Vz)
luminosity classification based on the relatively deep \specline{He}{2}{4686}
line also implies a young star early in its evolution \citep{PGMW92, WP92,
WB97}.  Stars \#21 and \#23 are both classified as O9.7~Ib, and stars \#25 and
\#30, the other visually bright stars in the western region, are both late type
stars (spectra were too noisy to classify).  Star 254 is clearly a foreground
star, with a negligible radial velocity and a very strong
\specline{Ca}{2}{3933} feature.  Star 263 was too noisy to classify, but it has
very strong Balmer lines, with a velocity consistent with being a Cloud member
($v \sim 210$--260~km~s$^{-1}$).

These spectra, and the photometry from the sources discussed above, form the
datasets with which we analyze in the following sections the stellar
populations of the region.


\section{COMPARISON AND ANALYSIS OF OPTICAL DATA } \label{sec:Comps}

From the three optical datasets, we selected those stars that are all within a
common, overlapping area.  The CCD fields are indicated by the solid,
rectangular outlines surrounding LH~2 in Figure~\ref{fig:uit_dss}b, and the
overlapping region has an area of $\sim 2.7 \times 2.6~{\rm arcmin}^{2} \sim
1466~{\rm pc}^2$.  In this area there are 222 stars from the CTIO85 catalog,
387 stars from the \citet{O96} catalog, 667 stars from the MCPS catalog, and 50
stars from the \UIT\ catalog.  These differences in the number of stars are
primarily due to different exposure times and limiting magnitudes, but a
detailed comparison also shows that in some cases there were also differences
in identification of close multiples, e.g., objects that were identified as
multiple stars in one dataset were identified as only one star in another.

\begin{figure*}

\epsscale{1.5}
\plotone{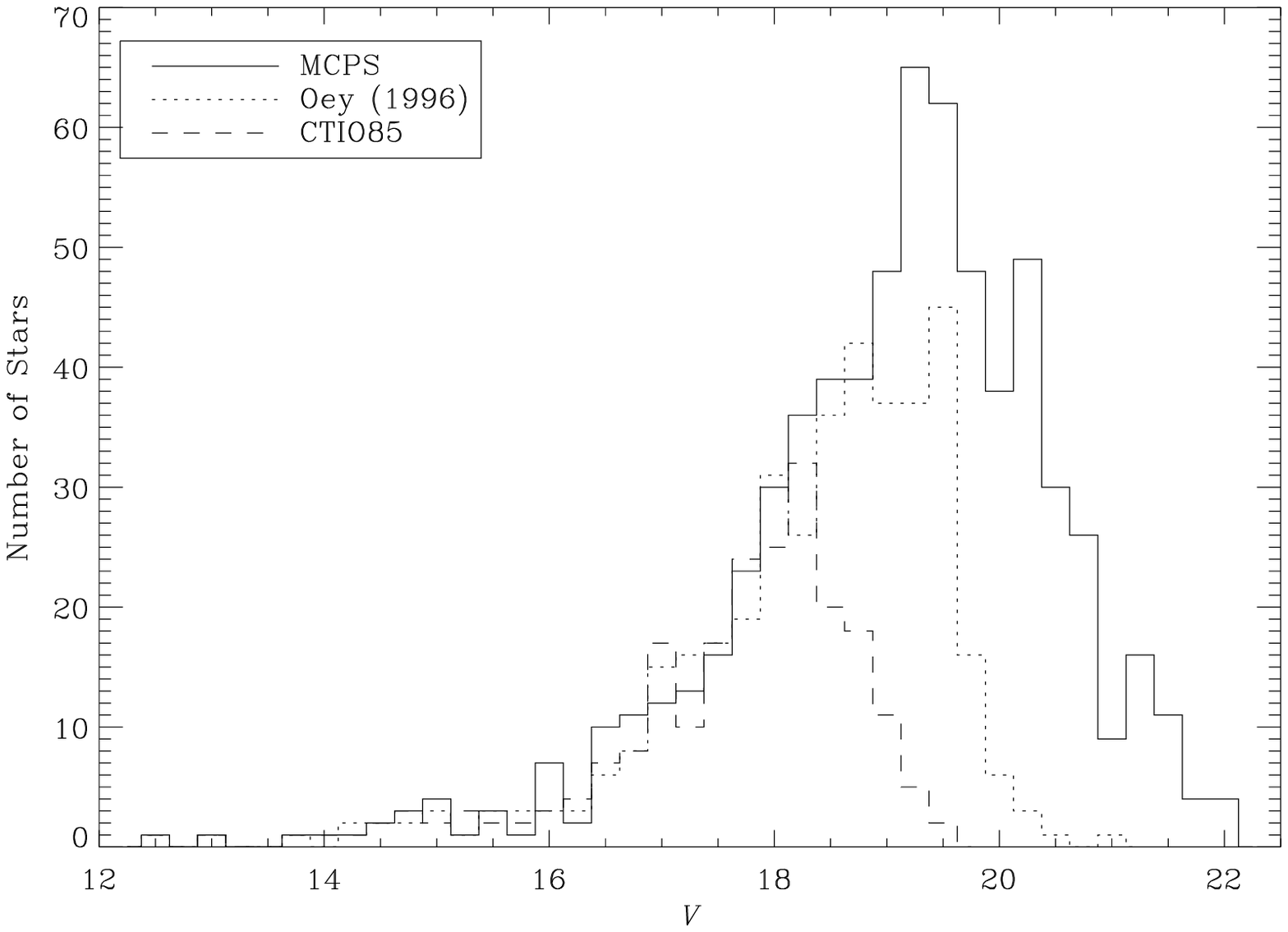}

\caption[]{A comparison of the $V$-filter observed magnitude distributions of
the three ground-based datasets.}
\label{fig:lf_comp}

\end{figure*}

Figure~\ref{fig:lf_comp} shows the magnitude distribution of the three optical
datasets.  The completeness magnitudes for CTIO85, \citet{O96}, and the MCPS
are roughly $V\sim 18$, 18.5, and 19.5~mag, respectively.  These magnitudes are
the values we will use in the following analysis and determination of the valid
mass range for the IMF calculations.


\begin{table*}
\caption{Comparison of $Q$ Indexes for Stars with Known Spectral Types \label{tab:Q}}

\begin{center}
\begin{tabular}{r l cccc}
\hline \hline
{Star} &
{Spectral Type} &
{$Q_{\rm Spec}$} &
{$Q_{\rm MCPS}$} &
{$Q_{\rm CTIO85}$} &
{$Q_{\rm O96}$} \\
\hline
  21 & O9.7 Ib      & $-1.11$ & $-0.64 \pm 0.12$ & $-0.81 \pm 0.04$ & \nodata \\
  23 & O9.7 Ib      & $-1.11$ & $-0.77 \pm 0.08$ & $-0.91 \pm 0.11$ & \nodata \\
  81 & O6 V         & $-0.95$ & $-1.09 \pm 0.11$ & $-0.77 \pm 0.04$ & $-0.93 \pm 0.02$ \\
  96 & O7.5 Vz      & $-0.93$ & $-0.63 \pm 0.09$ & $-0.75 \pm 0.03$ & $-0.82 \pm 0.01$ \\
 142 & O7 II:       & $-0.91$ & $-0.71 \pm 0.07$ & $-0.61 \pm 0.02$ & $-0.89 \pm 0.01$ \\
 180 & O9.5 V       & $-0.88$ & $-0.77 \pm 0.08$ & $-0.76 \pm 0.06$ & $-0.86 \pm 0.01$ \\
 223 & O8 II((f))   & $-0.91$ & $-0.77 \pm 0.61$ & $-0.60 \pm 0.01$ & $-0.87 \pm 0.01$ \\
 229 & O9.5 V:      & $-0.88$ & $-1.02 \pm 0.10$ & $-0.90 \pm 0.04$ & $-0.70 \pm 0.03$ \\
 250 & O7 V         & $-0.94$ & $-0.81 \pm 0.16$ & $-0.74 \pm 0.03$ & $-0.90 \pm 0.01$ \\
\cline{3-6}
 & & $\Delta Q = $ & $ 0.17 \pm 0.18$ & $ 0.27 \pm 0.07$ & $ 0.05 \pm 0.05$ \\
\hline \hline
\end{tabular}
\end{center}

\vspace*{-8ex}

\tablecomments{The uncertainties quoted for the individual $Q$ values are based
on the $UBV$ photometric uncertainties calculated by {\sc daophot}.  The
$\Delta Q$ values on the bottom row are $1/\sigma^2$-weighted averaged
differences in the sense: $Q_{\rm observed} - Q_{\rm Spec}$ (e.g., for $\Delta
Q > 0$, the observed colors are {\em redder\/} than the calibrated colors).
The uncertainties in $\Delta Q$ are similarly weighted standard deviations of
the mean.}

\end{table*}

\begin{figure*}

\plotone{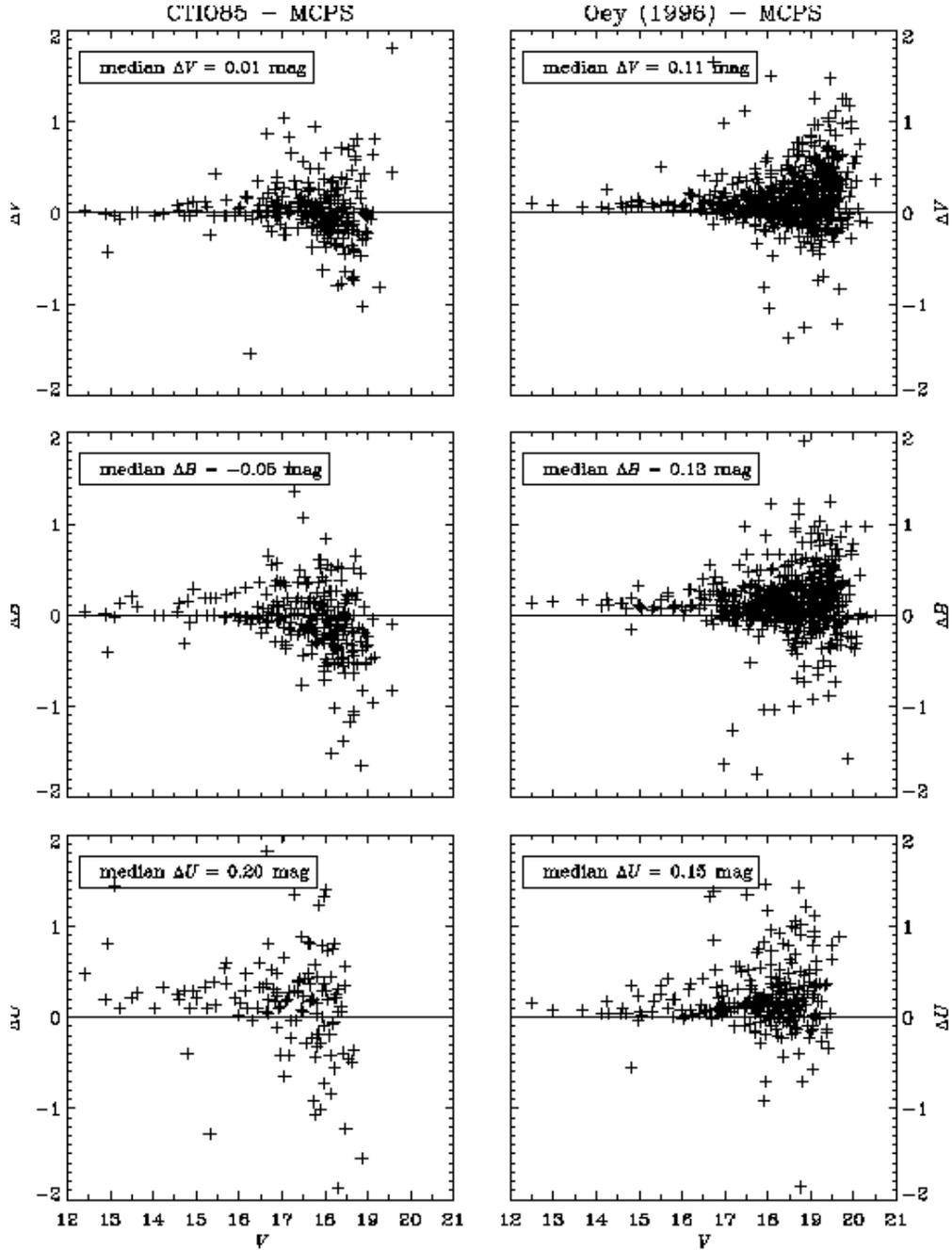}

\caption[]{Left column: A comparison of the photometry from the MCPS and
unpublished photometry from CTIO85 in the sense of CTIO85~$-$~MCPS.  Right
column: A similar comparison between photometry from the MCPS and \citet{O96}.}
\label{fig:comp_phot}

\end{figure*}

Figure~\ref{fig:comp_phot} shows a comparison of the magnitudes of stars in the
LH~2 region from the MCPS data and results from CTIO85 and \citet{O96}.  The
$V$ and $B$ data from CTIO85 are in reasonable agreement (within a few percent)
with the MCPS data, though there is a significant, 20\% zeropoint offset in the
$U$ photometry.  A comparison with the \citet{O96} data shows comparable
scatter, but 10--15\% offsets in all three filters.  For an independent
comparison of the photometry of the three datasets, in Table~\ref{tab:Q} we
compare the observed colors to the spectroscopically-calibrated colors of the
early type stars for which we have classified spectra
(Figure~\ref{fig:spectra}).  The \citet{O96} data show a good agreement on
average with the spectroscopic calibrations, and the MCPS and CTIO85 data show
offsets of about 20\% and 30\%, respectively, with fairly large scatter.  This
is consistent with the comparisons shown in Figure~\ref{fig:comp_phot}.  Such
large, systematic differences cannot be due to, e.g., incorrect values for the
reddening slope or intrinsic colors for LMC stars, since the same values were
used for all three datasets.  A probable culprit could be the aperture
corrections that were adopted in the original reduction of each dataset; the
corrections likely are the largest potential source of error in the photometric
calibrations.


To calculate effective temperatures (\Teff), bolometric corrections, bolometric
magnitudes (\Mbol), masses, and IMFs, we follow the general method used by
\citet{MSGD89} and numerous subsequent studies of OB associations in the
Magellanic Clouds \citep[e.g.,][]{MPG89, MLDG95, PGMW92, PG93, Wetal99}.  In
outline: first, a typical reddening, $E(B-V)_t$, is determined for the
association.  Then, for ``blue'' stars [stars having $-0.96 < Q < -0.40$, or
with both $-0.40 < Q < 0$ and $(B-V)_0 = (B-V) - E(B-V)_t < 0$], the \Teff\ is
calculated from their $Q$ index, and for other stars from their intrinsic
$(B-V)_0$ colors.  The bolometric correction is calculated from the \Teff, and
the resulting \Teff-\Mbol\ values are plotted on the Hertzsprung-Russell
diagram (HRD).  We use the Geneva evolutionary models \citep{SMMS93, Metal94}
with a metallicity of $Z=0.008$ to then bin the stars on the HRD by mass to
determine the ZAMS mass for each star, and thereby calculate an IMF.  The next
few paragraphs describe our particular steps in more detail.

To estimate $E(B-V)_t$ for the LH~2 region, we used three methods for each
dataset: ({\em i}) we estimated by eye the reddening necessary to ``slide'' the
observed color-magnitude diagrams to the intrinsic ZAMS color-magnitude
relationship; ({\em ii}) we use the $Q$ index for various subsets of the bluest
stars to calculate the typical reddening; and ({\em iii}) we compared the
observed and intrinsic $B-V$ colors for the stars that have spectroscopic
classifications.  All the datasets were in reasonable agreement, giving an
average value of $E(B-V)_t=0.20 \pm 0.02$ s.d.m., and a median value of
$E(B-V)_t=0.16$ (although the standard deviation of the mean is small, the
range of $E(B-V)$ values was quite large, with standard deviations typically
$\pm 0.15$, most likely due to intrinsic reddening variations within the
field).  This value is slightly larger than but consistent with results from
other analyses \citep{L72, L74, Y98}; those studies also find that LH~2 has an
extinction larger than typical for LMC associations.

For the special cases of stars with known spectral types, the reddenings were
determined from the intrinsic colors of \citet{J66}, and their \Teff\ were
determined using the calibrations of \citet{VGS96} and \citet{HM84}.  For other
``blue'' stars ($Q < -0.2$), the reddening and \Teff\ were calculated from
their $Q$ index.  For the remaining stars, the median typical reddening value
of $E(B-V)_t=0.16$ was used, and \Teff\ was calculated from $(B-V)_0$.  For all
stars, the bolometric correction was calculated from \Teff.


\begin{figure*}
\epsscale{1.2}
\plotone{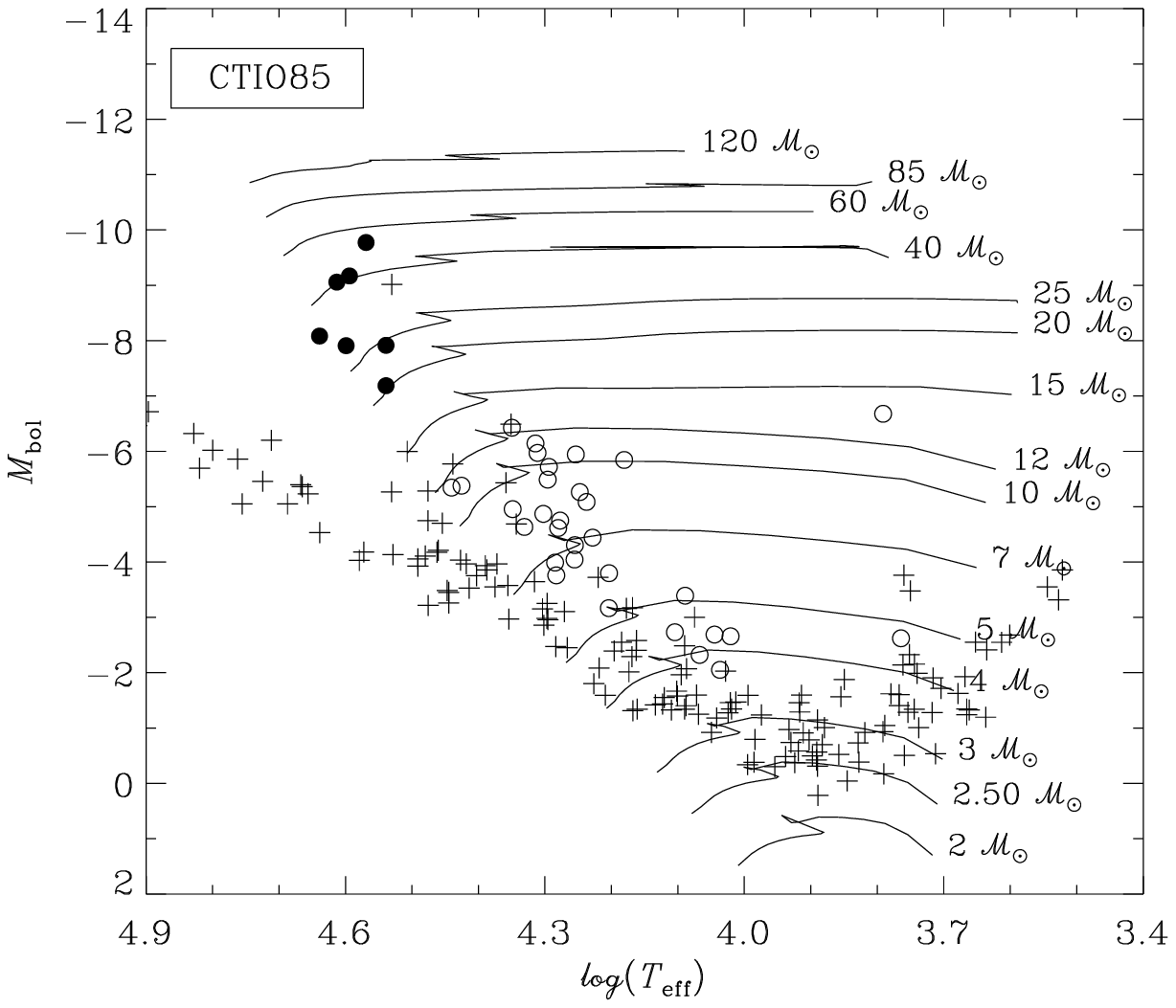} \hspace*{-10em} \ \ \ \ \ 
\plotone{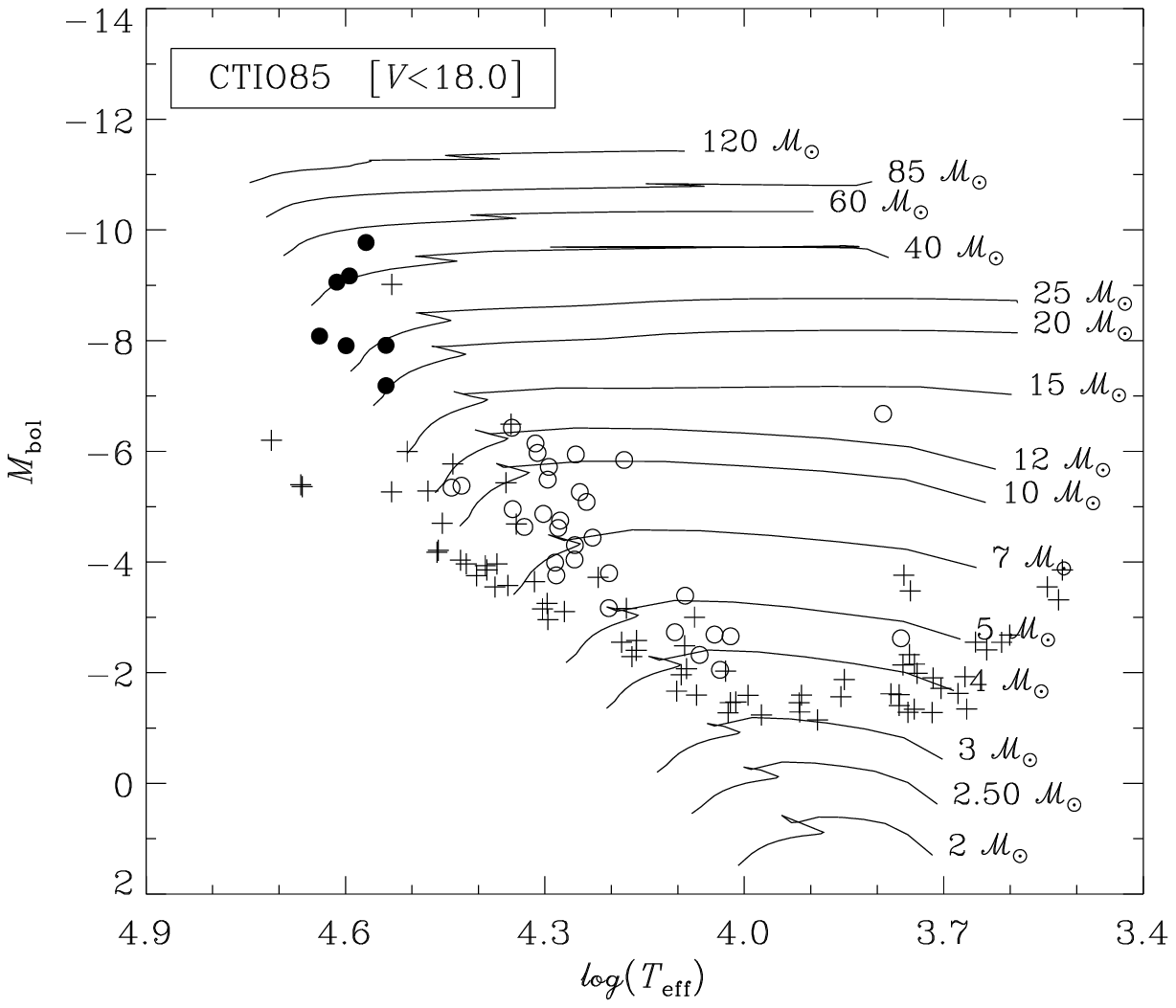}

\plotone{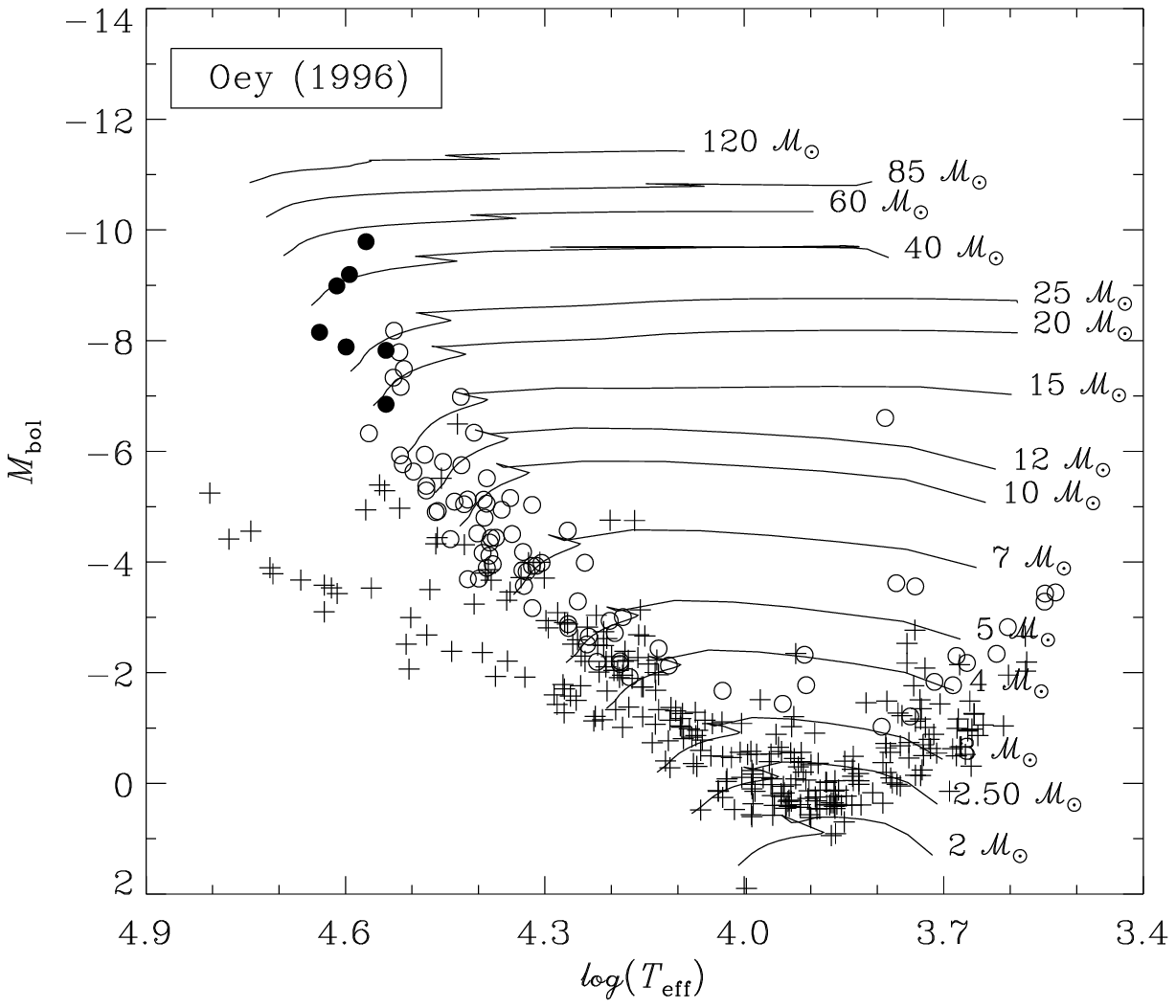} \hspace*{-10em} \ \ \ \ \ 
\plotone{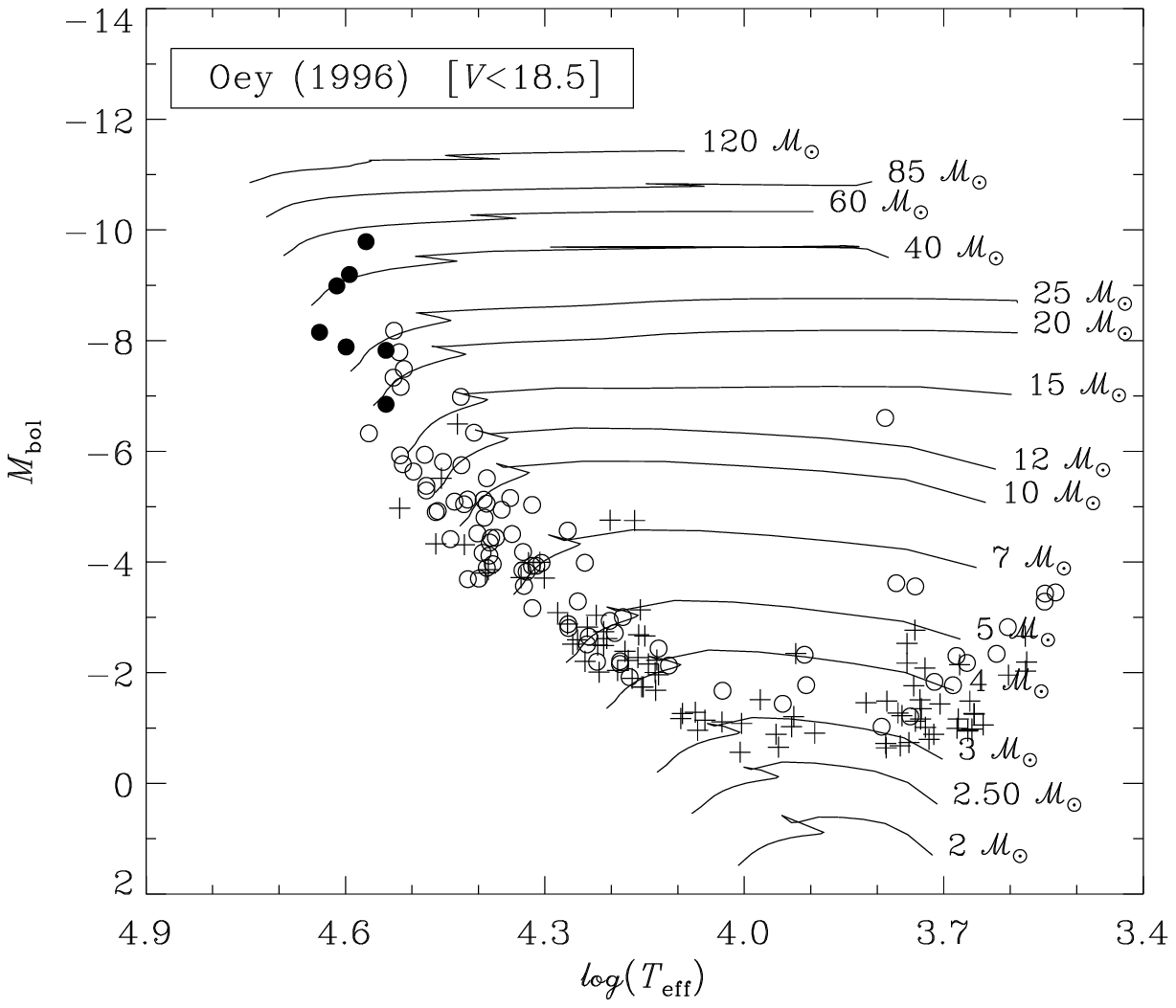}

\plotone{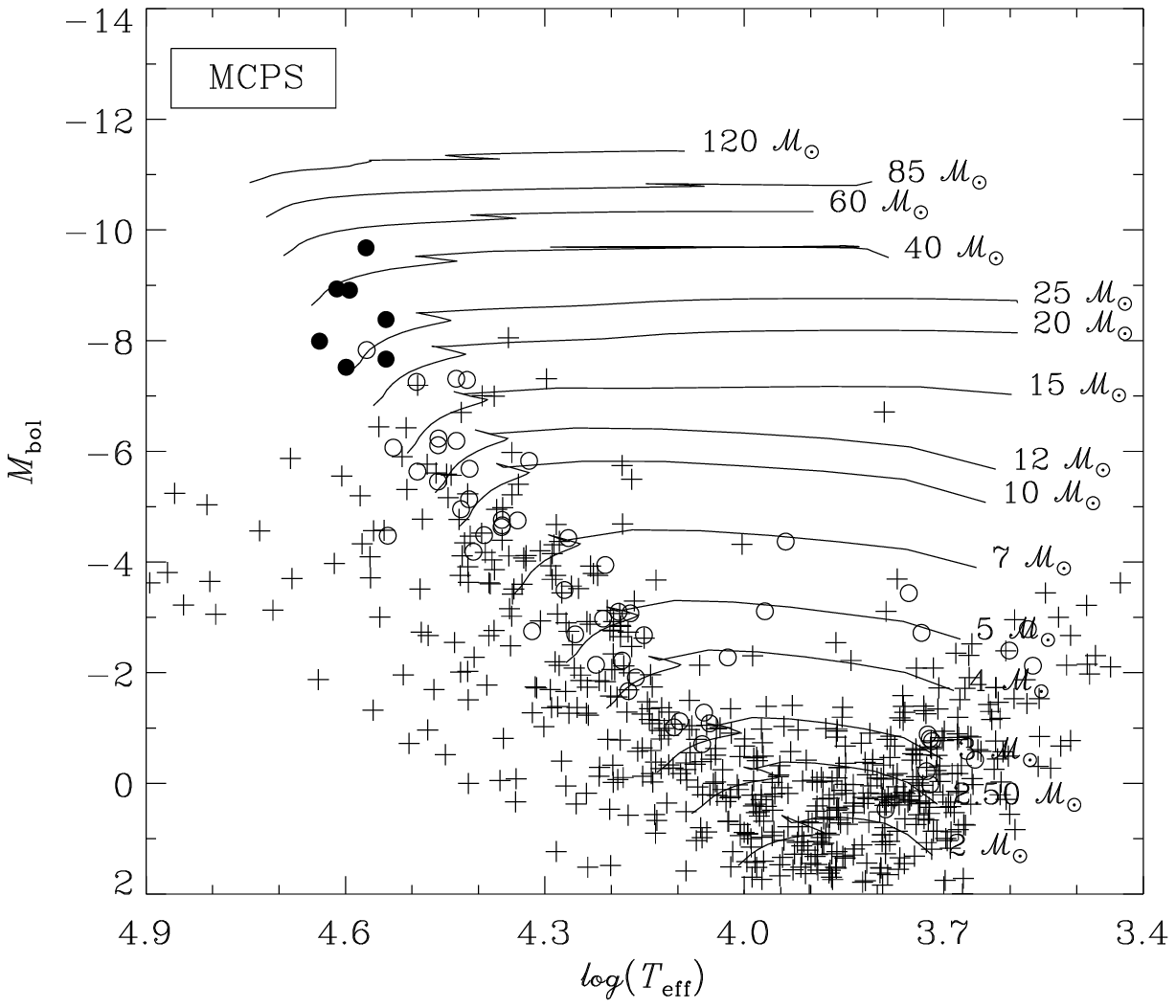} \hspace*{-10em} \ \ \ \ \ 
\plotone{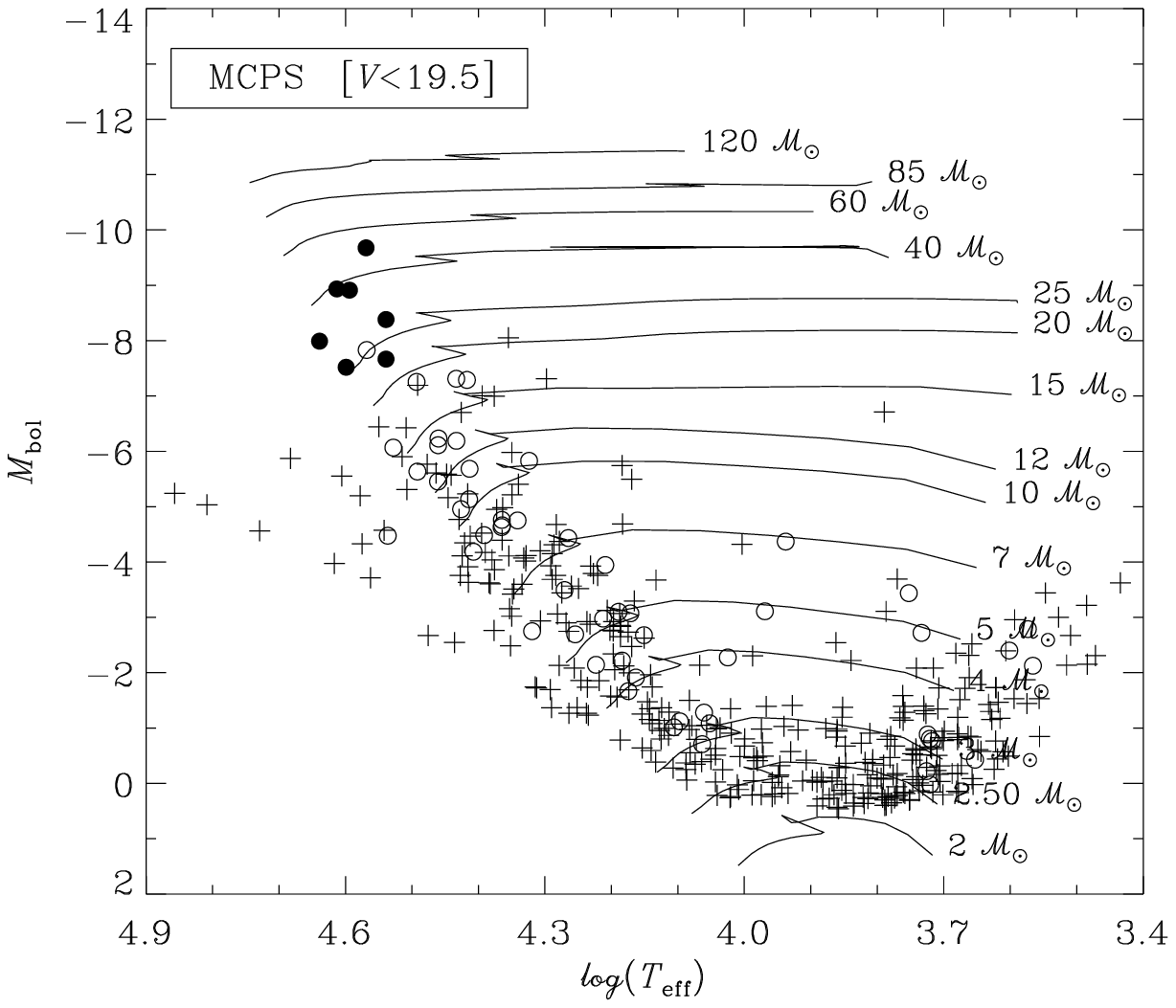}

\caption[]{The theoretical HRDs for the three ground-based datasets.  The
symbols indicate: stars with determined spectral types (filled circle,
$\bullet$), stars with photometry only and all uncertainties in $UBV <
0.07$~mag (open circles, $\circ$), and stars with photometry only and at least
one uncertainty $> 0.07$~mag (crosses, +).  The ``blue plume'' of stars to the
left of the ZAMS is an artifact of large errors for the fainter stars, as
discussed in the text.}
\label{fig:HRDs}

\end{figure*}

The resulting \Mbol\ and \Teff\ values for each dataset were plotted on the
HRD.  Figure~\ref{fig:HRDs} shows the HRDs for the data, overplotted with the
evolutionary tracks of \citet{SMMS93} for Z~=~0.008.  Stars with determined
spectral types are plotted with filled circles, stars with photometry only and
all uncertainties in $UBV < 0.07$~mag are plotted with open circles, and stars
with an uncertainty of $> 0.07$~mag in at least one filter are plotted with
crosses.  The ``blue plume'' of stars to the left of the ZAMS is a feature seen
in nearly all previous similar studies.  This feature is primarily due to faint
stars with large uncertainties; in particular, erroneously too-bright $U$
and/or $B$ magnitudes result in incorrectly large calculated \Teff\ and
bolometric corrections.  As seen when comparing the plots in
Figure~\ref{fig:HRDs}, these stars mostly disappear once a cutoff magnitude
consistent with the completeness magnitudes is used.  Only the MCPS dataset
still has a significant number of blue plume stars, perhaps indicating larger
photometric errors relative to the magnitude limit.


\begin{figure*}[ht]

\vspace*{5ex}

\epsscale{1.0}
\plotone{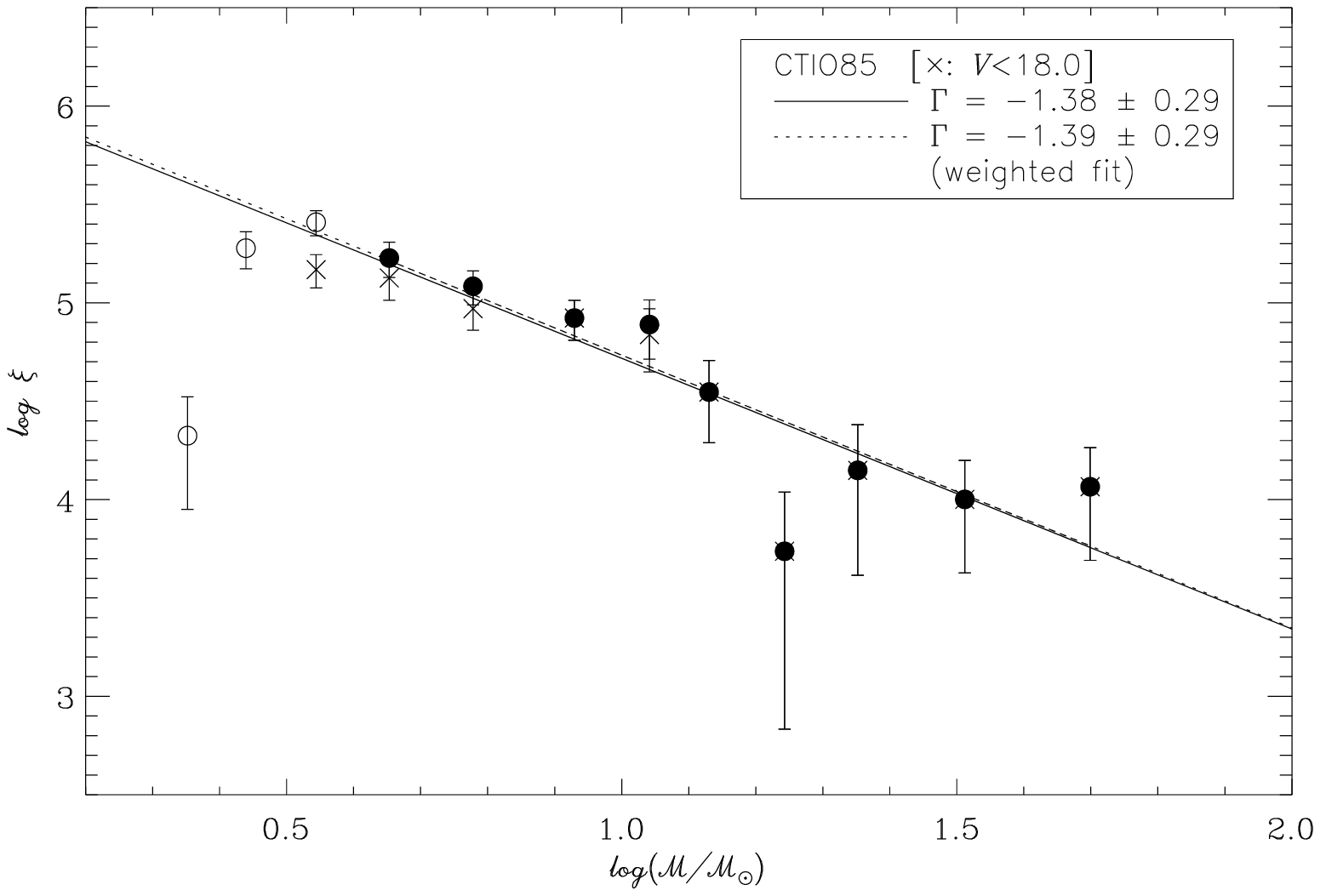}
\plotone{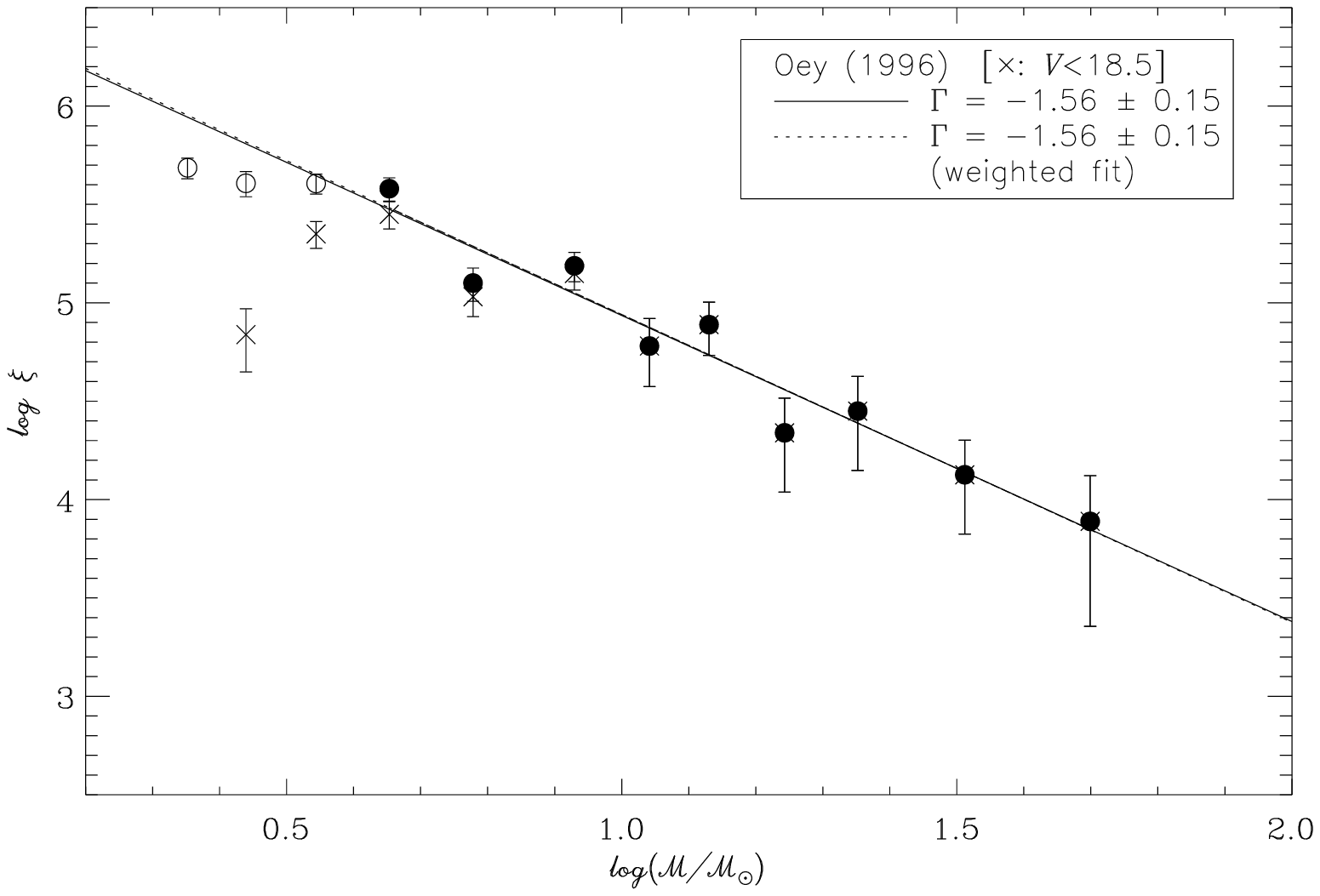}

\plotone{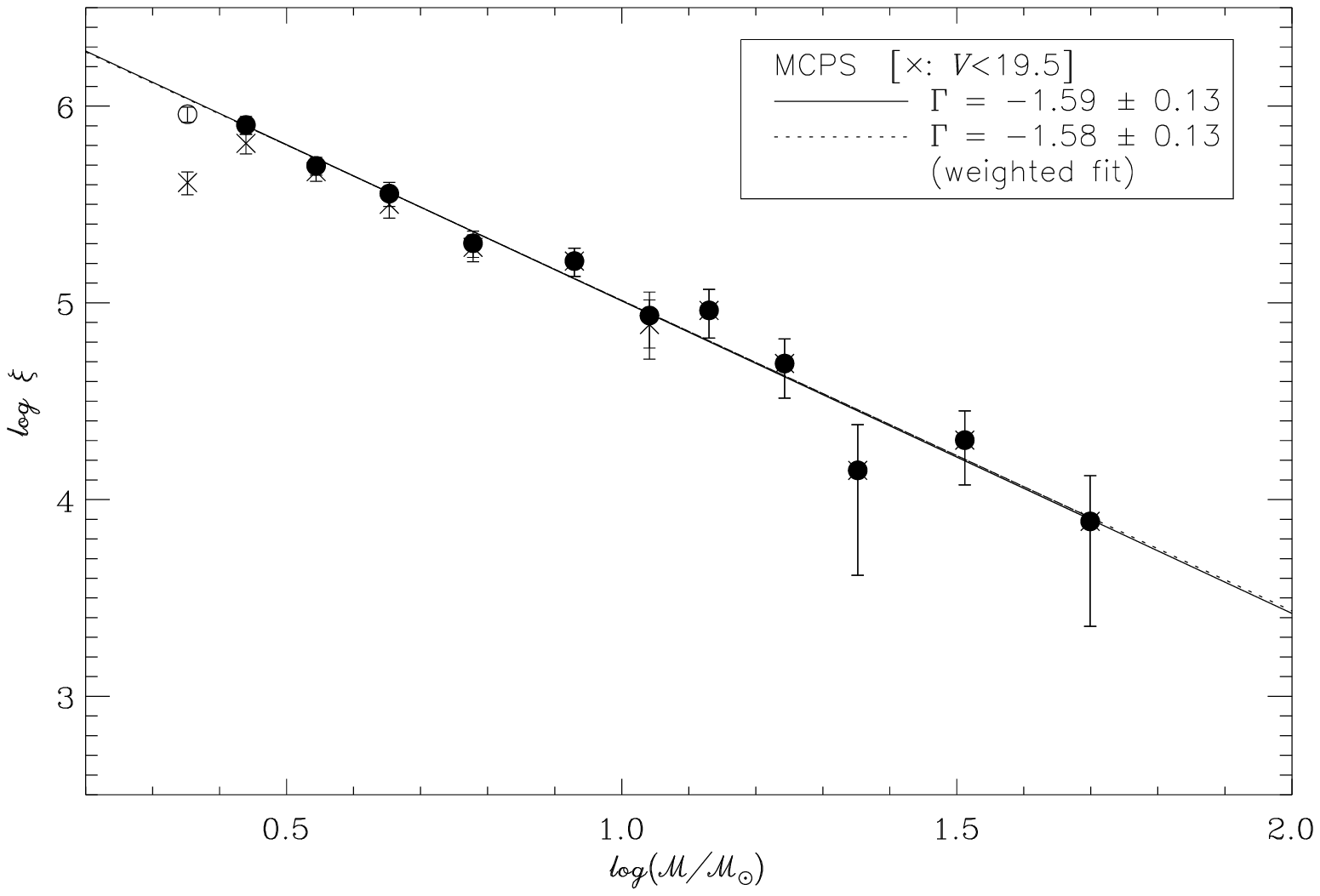}
\plotone{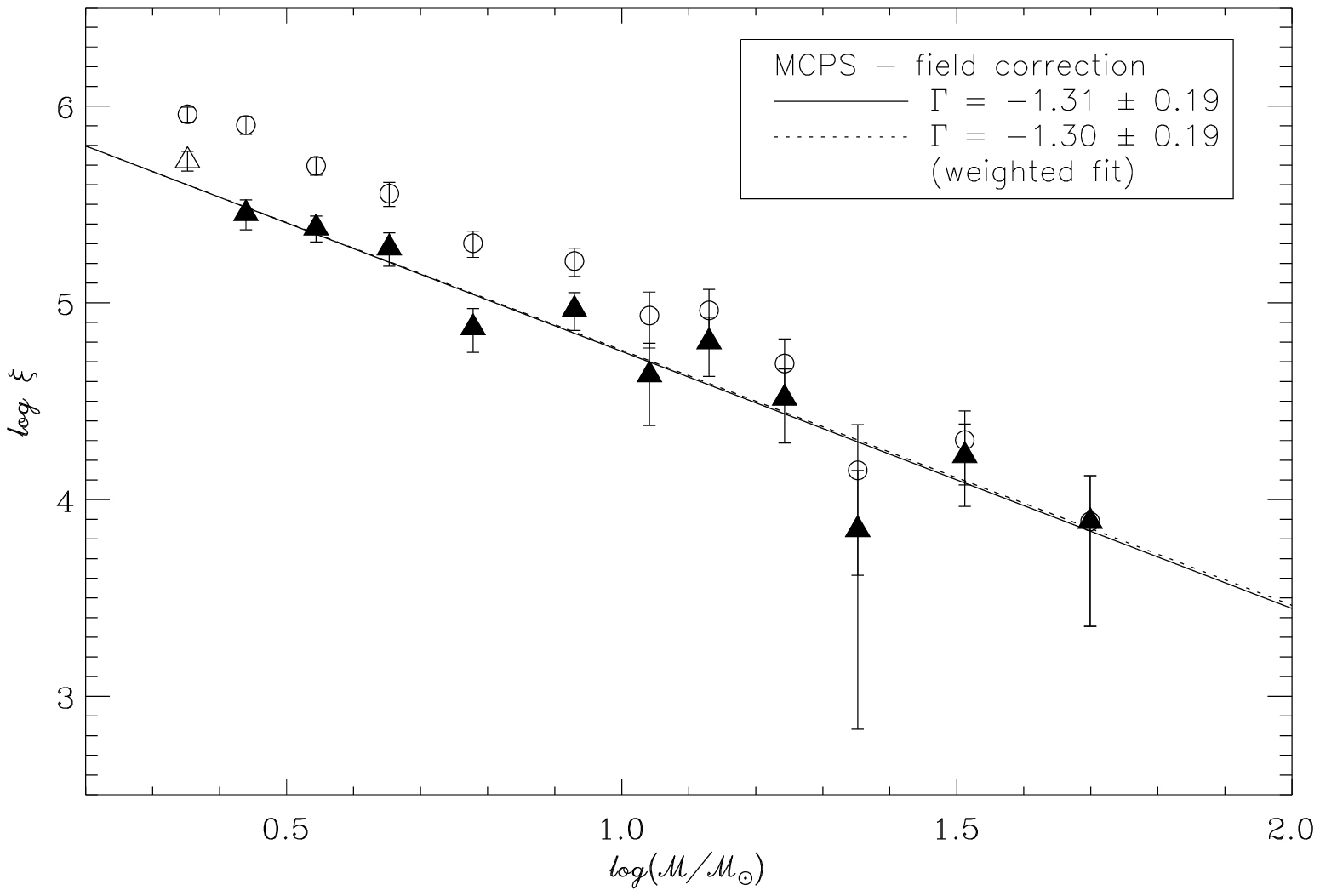}

\caption[]{The IMFs for the three ground-based datasets: a) CTIO85, b)
\citet{O96}, c) MCPS, d) MCPS after field star correction.  In all cases, the
solid line indicates an unweighted fit to the data points indicated by the
filled symbols, and the dashed line indicates a fit weighted by the number of
stars in each bin (in most cases, the fits indicated by the dashed and solid
lines are nearly indistinguishable).  The circles indicate all stars in each
dataset.  In panels~a--c, the crosses ($\times$) indicate those stars that are
brighter than the completeness magnitude cutoff determined for each dataset
(c.f.  Figure~\protect\ref{fig:lf_comp}).  For panel~d, the triangles indicate
the MCPS data for LH~2 after the estimated field star contribution has been
removed.}
\label{fig:IMFs}

\vspace*{5ex}

\end{figure*}

The resulting IMFs for the data are shown in Figure~\ref{fig:IMFs}.  These IMFs
include all the stars to the right of the ZAMS shown in the HRDs in
Figure~\ref{fig:HRDs}.  We have not included all the extant stars in the blue
plume in the IMF calculation since their true locations in the HRD are unknown
but are likely to be in the lower mass bins; however, to allow for some real
uncertainty in \Teff, we do include in the IMF those stars with \Teff\ up to
0.05~dex to the left of the ZAMS.  Also, no correction has been made for the
evolved stars (e.g, the star between the 12 and 15~\Msun\ mass tracks near
$\log \Teff \sim 3.8$), which are few in number so they do not have a
significant effect on the IMF slope.  The fairly narrow main sequence in the
HRDs implies that our assumption of coevality of this association is
reasonable.

The IMF slopes for the full datasets (panels a--c in Figure~\ref{fig:IMFs}) are
in remarkable agreement, in spite of the differences seen in the photometry and
the appearance of the HRDs.  However, we would like to emphasize two points:
all these IMFs rely on the same spectroscopic classifications for the bluest,
most massive stars, which reduces any effects from photometric differences; and
clearly it is possible to get similar or even ``correct'' IMF {\em slopes\/}
fortuitously with data that have large uncertainties.  \citet{M98} points out
that a systematic error in \Mbol\ (due to, e.g., errors in the photometry,
calibrations, or adopted distance modulus) does not change the slope of the
IMF; the stars tend to shift uniformly from one bin to another without
necessarily changing the slope of the IMF.

\begin{table*}[ht]
\caption{Stars with Spectroscopic Observations \label{tab:spec}}

\begin{center}
\begin{tabular}{r rr l rrr c rrrr}
\hline \hline
\multicolumn{4}{c}{} &
\multicolumn{3}{c}{$T_{\rm eff}$ (K)} &
{} &
\multicolumn{4}{c}{$E(B-V)$} \\
\cline{5-7} \cline{9-12}
{Star} &
\multicolumn{2}{c}{$\alpha$ \ \ \ (2000.0) \ \ \ $\delta$} &
{Spectral Type} &
{spec\tablenotemark{a}} &
{phot$_Q$\tablenotemark{b}} &
{phot\tablenotemark{c}} &
{} &
{spec\tablenotemark{a}} &
{phot$_Q$\tablenotemark{b}} &
{phot\tablenotemark{c}} &
{match\tablenotemark{d}} \\
\hline
  21 &   4:51:45.98 & --69:20:25.9 & O9.7 Ib      & 30500 & 17000 & 23000 & &   0.01 & --0.04 &   0.10 &   0.19 \\
  23 &   4:51:47.05 & --69:19:04.3 & O9.7 Ib      & 30500 & 22000 & 27000 & &   0.10 &   0.09 &   0.19 &   0.21 \\
  25 &   4:51:47.39 & --69:19:24.5 & late type    & \nodata & \nodata & \nodata & & \nodata & \nodata & \nodata & \nodata  \\
  30 &   4:51:48.55 & --69:19:03.3 & late type    & \nodata & \nodata & \nodata & & \nodata & \nodata & \nodata & \nodata  \\
  81 &   4:52:00.66 & --69:20:49.7 & O6 V         & 43560 & 50000 & 42500 & &   0.28 &   0.32 &   0.28 &   0.28 \\
  96 &   4:52:02.40 & --69:20:33.3 & O7.5 Vz      & 39730 & 17000 & 26000 & & --0.02 & --0.11 &   0.07 &   0.17 \\
 142 &   4:52:09.12 & --69:20:35.7 & O7 II:       & 39290 & 20000 & 26000 & &   0.06 & --0.01 &   0.10 &   0.20 \\
 180 &   4:52:11.88 & --69:20:25.2 & O9.5 V       & 34620 & 26000 & 30000 & &   0.08 &   0.05 &   0.11 &   0.13 \\
 223 &   4:52:18.22 & --69:20:11.7 & O8 II((f))   & 37090 & 22000 & 19000 & &   0.13 &   0.09 &   0.01 &   0.22 \\
 229 &   4:52:19.32 & --69:20:49.3 & O9.5 V:      & 34620 & 50000 & 24000 & &   0.46 &   0.51 &   0.07 &   0.17 \\
 250 &   4:52:21.34 & --69:20:29.4 & O7 V         & 41010 & 28000 & 40000 & &   0.17 &   0.13 &   0.21 &   0.22 \\
 254 &   4:52:22.26 & --69:20:18.6 & foreground   & \nodata & \nodata & \nodata & & \nodata & \nodata & \nodata & \nodata  \\
 263 &   4:52:23.96 & --69:19:58.9 & LMC member   & \nodata & \nodata & \nodata & & \nodata & \nodata & \nodata & \nodata  \\
\hline \hline
\end{tabular}
\end{center}

\vspace*{-8ex}

\tablecomments{The few values of $E(B-V)<0$ are probably due to photometric
errors (c.f., Table~\protect\ref{tab:Q}) in cases of low intrinsic
reddening.  It is possible that for star \#96 another factor (e.g.,
multiplicity) also may be responsible.}

\tablenotetext{a}{$T_{\rm eff}$/spec is the effective temperature as a function
of spectral type using the calibration of \citet{VGS96}, and $E(B-V)$/spec is
the reddening calculated from the the observed $(B-V)$ color minus the
intrinsic $(B-V)_0$ color based on the spectral type.}

\tablenotetext{b}{$T_{\rm eff}$/phot$_Q$ and $E(B-V)$/phot$_Q$ are the
effective temperature and reddening determined by fitting the UV and optical
data using a fixed $E(B-V)$ calculated from the $Q$ index based on the observed
$UBV$ photometry.}

\tablenotetext{c}{$T_{\rm eff}$/phot and $E(B-V)$/phot are the effective
temperature and reddening determined by fitting the UV and optical data
allowing $E(B-V)$ to be a variable parameter of the fit.}

\tablenotetext{d}{$E(B-V)$/match is the reddening required to get the fitted
$T_{\rm eff}$ to match the $T_{\rm eff}$/spec.}

\end{table*}

The MCPS data have the advantage that they cover sufficient area to allow
correction for contamination of background stars (i.e., LMC field stars) in the
area of LH~2.  We selected two large fields (indicated by the dashed-lined
boxes in Figure~\ref{fig:uit_dss}b) by eye to serve as background fields.
Stars in these fields were run through the same calculations to determine IMF
masses.  The number of stars in each mass bin was scaled by the relative sizes
of the LH~2 and the background fields, and then was subtracted from the IMF
plotted in Figure~\ref{fig:IMFs}c.  The resulting field star corrected IMF is
shown in Figure~\ref{fig:IMFs}d.  This IMF is notably flatter than the
uncorrected IMF.\footnote{This result does {\em not\/} imply that the field
star IMF is steeper than the OB association IMF.  To correctly calculate the
field star IMF, one must include the stellar lifetimes and assumed star
formation history.  All we have done here is to correct for field stars that
appear within the boundaries of the OB associations, and which were treated as
coeval association members.}  The fact that field stars can affect the observed
IMF is not a new idea, but it is a correction that is often not made, possibly
because most observations do not cover a large enough area to make the
measurement and correction and it is assumed that there are not many or
sufficient OB-type field stars to bias the sample.  Our result simply
emphasizes the fact that this correction {\em is\/} important even for OB-type
stars, and, if the IMF slope is constant and universal, such field
contamination could possibly even be a major factor causing the wide range of
observed IMF slopes.


\section{ANALYSIS OF \UIT\ AND MCPS DATA } \label{sec:UV}

\subsection{Determination of Reddening and Effective Temperatures}

The \UIT\ data were combined with the MCPS data using all MCPS stars within the
field-of-view of the \UIT\ image, which contains 80733 stars from the MCPS
catalog and 3533 stars in the \UIT\ catalog.  Because the MCPS data do not
extend across the entire declination range of the \UIT\ field (see Figure
~\ref{fig:uit_dss}b), we will only use the overlapping sub-region for our
analysis in this paper.  In that region, there are 2770 MCPS stars that have UV
photometry from the \UIT\ catalog.  These stars were matched by positional
coincidence with a 3~arcsec tolerance.

To compare our photometry to \citet{K92,K93} models, we use UV and optical
magnitudes derived from filter functions convolved with Kurucz spectra.  The
UV magnitudes were derived by Landsman (personal communication) by convolving
Kurucz spectra with the \UIT\ response curve.  Optical/IR standard broad band
colors and magnitudes for the Kurucz models were calculated by \citet{BCP98};
however, those data were determined for solar abundances.  From those authors
we obtained their data for models with $[M/H]=-0.5$, more appropriate for LMC
metallicity \citep{KKM93}.


\begin{figure*}[ht]

\epsscale{1.5}
\plotone{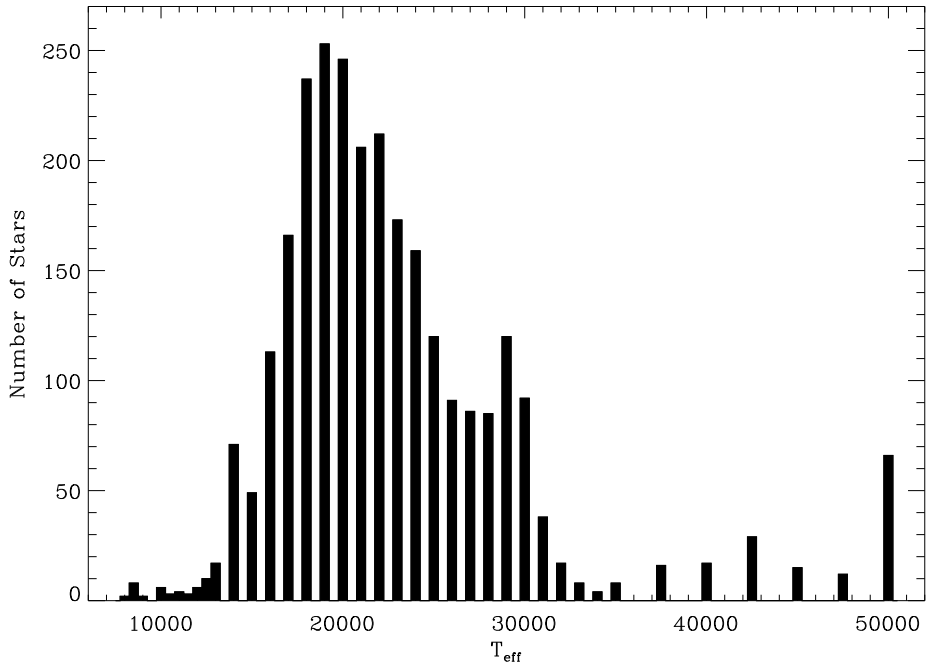}

\caption[]{The number distribution of effective temperatures (\Teff) determined
by fitting the UV and optical photometric data and allowing the $E(B-V)$
reddening to be a variable of the fit for each star.}
\label{fig:teff_dist}

\end{figure*}

The observed magnitudes were compared to the synthetic photometry of the models
to determine the best matching model (\Teff\ and surface gravity) using a
number of different programs and best-fit estimators.  In Table~\ref{tab:spec}
we show the results of some of these fits for those stars that have
spectroscopic observations, and therefore, good estimates of their true \Teff.
In one case (the ``phot$_Q$'' columns), we first calculated the $E(B-V)$ for
each star using the $Q$ index as determined from the $UBV$ photometry
(Table~\ref{tab:Q}), dereddened the observed colors, then performed the fit of
the data to the models.  In another case (the ``phot'' columns) we allowed
$E(B-V)$ to be a free parameter in the fit.  For comparison, in the ``spec''
columns, we also show the $E(B-V)$ calculated from the observed $B-V$ color and
the intrinsic $(B-V)_0$ color based on the color~vs.~spectral type
calibration.  In all cases, for the reddening law we use the functional form of
\citet{FM86, FM88} with the ``average LMC'' coefficients derived by
\citet{MCG99}, and fits were weighted using the {\sc daophot}-derived
photometric errors.  (In many cases for the COTS, the $I$ data were saturated,
which is reflected in the quoted errors.)

In nearly all cases, the \Teff\ calculated in the fit underestimates the
``true'' \Teff\ as determined from calibration with spectral type
\citep{VGS96}, though about half have reasonably good fits (i.e., the
difference in \Teff\ is $<0.1$~dex).  In the ``match'' column of
Table~\ref{tab:spec} we show the $E(B-V)$ that would be required to get the
fitted \Teff\ to match the spectroscopic \Teff.  If the calculated $Q$ is too
red, the resulting $E(B-V)$ and \Teff\ values will also be too low.  If we
adjust the MCPS data by the value of $\Delta Q = 0.17$ shown in
Table~\ref{tab:Q}, the fitted \Teff\ values are significantly improved, but
still too low on average.  It is probably the case that this systematically
too-low \Teff\ is not only a factor of an incorrectly derived reddening, but
possibly also due to a model calibration error; the fits are most sensitive to
the UV--optical colors, which involve two different photometric systems that
may not be well cross-calibrated.  However, we should point out that if we use
only the optical data to calculate $E(B-V)$ and/or perform the fit, the fitted
\Teff\ values even further underestimate the true temperatures in most cases,
so the addition of the UV data, while not a sufficient replacement for getting
spectral types, definitely improves our estimate of \Teff\ for O-type stars.
So although the inclusion of the UV data does not allow us to definitively
identify the O-type stars and determine their \Teff, it allows us to establish
a better lower limit to the number of O-type stars in the catalog than optical
data would alone.
\footnote{If these fitted \Teff\ values were systematic in their differences
from the spectroscopic values, we perhaps could apply a correction.  However,
there are too few stars in Table~\protect\ref{tab:spec} to determine a reliable
correction factor.  We are in the process of performing larger-scale study of
Magellanic Cloud stars with spectra and UV+optical photometry to study this
effect, which could be due to many factors including errors or uncertainties
in: systematic photometry errors (see Table~\protect\ref{tab:Q}); applying a
Galactic \Teff~vs.~spectral type calibration to LMC stars; the Kurucz models
for LMC metallicity; the convolution of the filter function with the models to
obtain colors on the standard system; and/or the fitting method used to find
the best match of the model colors to the data.  It is also possible that the
uncertainty may not be correctable with this kind of data:  although one may
optimize the the calibration for blue stars, the \Teff\ for O-type stars can
increase significantly without any real, notable change in the observable
color.  This is an asymmetric effect, i.e., a {\em large increase\/} in
\Teff\ is harder to detect via optical (and even UV) colors than a {\em
relatively smaller decrease\/} in \Teff.  This would imply that, on average,
any calibration of \Teff~vs.~photometric color will tend to underestimate the
temperatures for O-type stars.  See also the discussion by \citet{M98} on how
this asymmetry can affect the slope of the IMF.}


\begin{figure*}[ht]

\epsscale{1.0}
\plotone{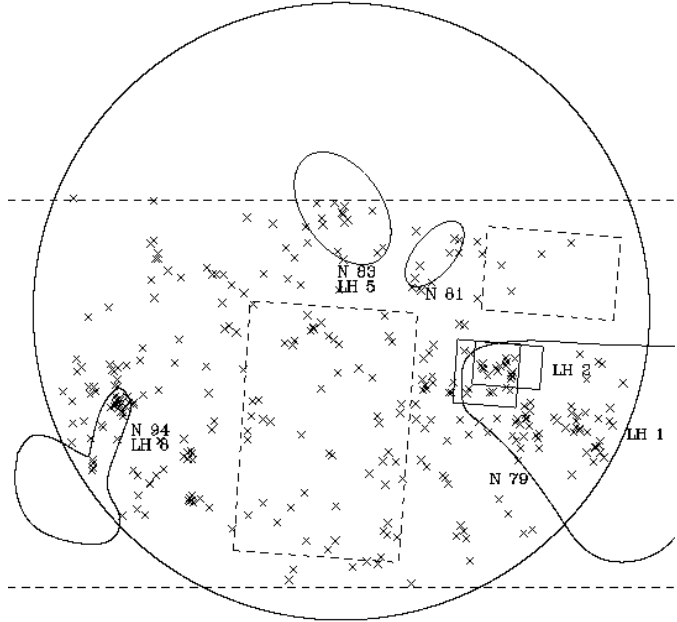}

\caption[]{The spatial distribution stars with $\Teff \ge 30,000$~K determined
by fitting the UV and optical photometric data as discussed in the text.  The
outlines indicate the same regions shown in Figure~\protect\ref{fig:uit_dss}b.}
\label{fig:teff30k}

\end{figure*}

We then performed these fits to all 2770 stars in our dataset that have UV and
optical data, and the temperature distribution is shown in
Figure~\ref{fig:teff_dist}.  The qualitative shape of the distribution is
roughly as expected: it rises up quickly at low \Teff\ (due to incompleteness
and observational bias against the fainter, cooler stars), then the numbers
steadily decrease going to higher \Teff, reflective of the IMF.  There are two
unusual features: a plateau (and small spike) between 26,000 and 30,000~K, and
a spike at \Teff\ = 50,000~K.  It is unclear if these are artifacts of the fit
or real features.  For example, the spike at 50,000~K could be because that is
the maximum \Teff\ in the models, so all hotter stars are put into that bin.
Also, another point of concern is that we find a slight trend of hotter stars
having larger $E(B-V)$ values on average (although there is a large scatter;
stars of all temperatures do show a wide range of reddenings).   Since the
reddening is a variable in the fit, if the reddening is overestimated for a
star, then the \Teff\ may be overestimated.  However, a similar though less
steep trend is seen if one compares the $E(B-V)$ derived purely from the $UBV$
photometry via the $Q$ index.  So this trend may be real, and may reflect the
situation that the hottest, most massive stars also are the shortest-lived, and
therefore are more likely to still be in or near their birth regions, which
will have larger than average extinction.  In fact, the $Q$-derived $E(B-V)$
values tend to be larger on average than the fitted values (contrary to the
trend seen with the few stars in Table~\ref{tab:spec}); if the reddening values
are, indeed, larger than the fitted values, then our \Teff\ estimates may be
too low, and there may be more stars with higher temperatures than shown in
Figure~\ref{fig:teff_dist}.

\subsection{Candidate O-type Stars}

Taking our results at face value, there are 322 stars with $\Teff \ge
30,000$~K, the typical temperature of the latest O-type star.  We note that the
comparison in Table~\ref{tab:spec} and the previous discussion about the
reddening imply that our fits may tend to underestimate the true temperatures.
In Table~\ref{tab:spec}, O-type stars have fitted temperatures as low as
$\sim$20,000~K.  We find 1820 stars with fitted $\Teff \ge 20,000$~K.  Given
that the area of the dataset is $\sim$1.4\xten{5}~pc$^2$, these numbers
translate to a density of one O-type star per $\sim$80--430~pc$^2$
(0.4--2.1~arcmin$^2$).  For comparison, the density in the solar neighborhood
is less one O-type star per 40,000~pc$^2$ \citep{GCC82}.

Figure~\ref{fig:teff30k} shows the spatial distribution of the stars with
$\Teff \ge 30,000$~K in the region we analyzed.  Note the large number of COTS
well-distributed outside of the boundaries of the classical OB associations:
about 200 of the hot stars, nearly two-thirds of the COTS, are outside the
boundaries of the ``N'' and ``LH'' regions shown in Figure~\ref{fig:teff30k}.
How many of these may be true field stars?  \citet{MLDG95} define ``field
stars'' as those that are farther from the boundary of any OB association than
the distance an O-type star could travel in its lifetime (10~Myr for late-type
O stars) at a typical dispersion velocity of $\sim 3$~km~s$^{-1}$ relative to
the parent molecular cloud.  This distance is about 30~pc, or 2~arcmin at the
distance of the LMC.  This is a reasonable definition for most cases.  If we
count only those stars beyond $\sim$2~arcmin from any of the boundaries, that
still leaves about 160 stars.  {\em Approximately equal numbers of COTS are
found in the field and in OB associations.\/}  This result is in accord with
earlier suggestions from the Galactic study of \citet{GCC82}; their Figure~2
shows that perhaps the majority of the O stars within 3~kpc of the Sun are not
in OB associations.  Our result can be extrapolated to the general distribution
of massive stars in the LMC only if one assumes that the region studied here is
large enough to provide a representative and proportional sample of field and
association areas.  The density of ``field'' OB stars in this region is roughly
one star per 700~pc$^2$ (3.5~arcmin$^2$).

Because of the uncertainties in obtaining \Teff\ from photometry alone, these
estimates will have to be verified with spectroscopic observations.  It is
unclear if our estimates of the O-type star population may be high or low: some
of these COTS will turn out to be later-type stars because we have
over-estimated their \Teff, but it is perhaps even more likely that there were
many O-type stars that were missed because we under-estimated their
\Teff\ (e.g., as in Table~\ref{tab:spec}, many O-type stars were fitted $\Teff
< 30,000$~K).  Also, because we have only considered those stars with both UV
and optical photometry, some O-type stars may have been missed in our analysis
because they were too heavily obscured to be detected in the UV, and this may
be a stronger effect in the younger star forming regions where extinction is
higher.  In light of these uncertainties, our analysis has been reasonably
conservative, and we conclude that our results provide a lower limit to the
total population of O-type stars in the field.


\section{SUMMARY}

In this paper we have analyzed the UV photometric data from \UIT\ along with
the optical ($UBVI$) data from the Magellanic Cloud Photometric Survey and
other ground-based sources to study the stellar content within and near N~79 in
the southwest region of the LMC. Our analysis of these data gives the following
results:

\begin{itemize}

\item Comparisons of three different optical datasets for the LH~2 OB
association show that although the datasets exhibit median photometric
differences of up to 30\%, the resulting observed, uncorrected IMFs are
reasonably similar, typically $\Gamma \sim -1.6$ in the 5--60~\Msun\ mass
range.

\item When we correct for the background contribution of field stars in the one
dataset where this is possible, the calculated IMF flattens to $\Gamma
= -1.3 \pm 0.2$ (similar to the Salpeter IMF slope).  This implies that the
background contribution of stars---even for massive stars---may be an important
factor to the range of IMF slopes found in the literature.

\item Fitting the UV+optical data to Kurucz models, we find 322 stars with
fitted $\Teff \ge 30,000$~K (the lower limit for \Teff\ for a typical, latest
type O star), and 1820 stars with fitted $\Teff \ge 20,000$~K (the lower limit
for the fitted \Teff\ for known O-type stars used for comparison in
Table~\ref{tab:spec}).

\item  We find evidence that the number of candidate O-type ``field'' stars is
roughly equal to the number of such stars in OB associations.  This
distribution is very interesting in that it provides strong confirmation of the
conventional wisdom; see \citet{GCC82}.

\end{itemize}

This work shows the potential of the large-scale analysis of stellar
populations that will be possible when the MCPS is complete and combined with
the \UIT\ dataset.  The results of this and similar studies are leading us to
make more detailed observations of the massive star content of the field region
in both Magellanic Clouds to better understand the origin of these stars and
their IMF.   We are presently conducting spectroscopic observations and further
analysis to determine if these are true isolated field stars, not born in OB
associations, or if they are high-velocity runaways or members of many
previously unrecognized, low-density OB associations.  We will also include
field regions of the LMC and SMC that are notably less active in present star
formation (e.g., farther from or containing fewer and smaller OB associations)
to see if our results here are representative of the global distribution of
massive stars.  Ultimately, these data will allow us to resolve discrepant
measurements of the field star IMF and the origins of massive stars.  If there
is a significant population of massive field stars that have formed in situ as
isolated events, it will have important implications on concepts and models of
star formation.

\acknowledgments

Thanks to F.~Castelli for making their data available to us, N.~Walborn for
independent checking of spectral types, and W.~Landsman for numerous valuable
discussions.  The Digitized Sky Surveys were produced at the Space Telescope
Science Institute under U.S. Government grant NAGW-2166 with supplemental
funding provided by the European Southern Observatory for sky-survey work. The
image used in this paper is based on photographic data obtained using the UK
Schmidt Telescope operated by the Royal Observatory Edinburgh, with funding
from the UK Particle Physics and Astronomy Research Council and the
Anglo-Australian Observatory.    Support for J.Wm.P. for this project was
through NASA LTSA grant NAG5-9248 and a subcontract from
\UIT\ through Raytheon.  Funding for the \UIT\ project has been through the
Spacelab Office at NASA Headquarters under project number 440-51.  D.Z.
acknowledges financial support from NSF grant AST~96-19576, NASA LTSA grant
NAG5-3501, a David and Lucile Packard Foundation Fellowship, and an Alfred P.
Sloan Fellowship.



\end{document}